\documentclass[11pt,a4paper]{article}
\usepackage[a4paper, left=1in, right=1in, top=1in, bottom=1in, includehead, includefoot, foot = 0pt, head=0pt]{geometry} 

\usepackage{amsthm,amsmath,natbib}
\RequirePackage[colorlinks,citecolor=blue,urlcolor=blue]{hyperref}

\usepackage{amsfonts}

\usepackage{tikz}											
\usetikzlibrary{shapes,arrows,decorations.pathmorphing,backgrounds,positioning,fit,petri}
\usepackage{pgfplots}									
\pgfplotsset{compat=newest}		
\usepackage{subcaption}									
\usepackage{array}
\usepackage{siunitx}									
\usepackage{enumitem}  

\newcommand{\bs}{\boldsymbol}
\def\IBNR{\mathrm{nr}}  
\def\E{\mathrm{E}} 

\def\r{\mathrm{r}}  
\def\P{\mathrm{P}} 
\def\W{\mathrm{W}}

\DeclareMathOperator{\dow}{\textsf{dow}}
\DeclareMathOperator{\wday}{\textsf{wday}}

\usepackage{xcolor}

\def\tcr{}
\usepackage{verbatim}

\usepackage{colortbl}
\usepackage{titling}

\usepackage{authblk} 

\begin{document}

\title{Modeling the occurrence of events subject to a reporting delay via an EM algorithm}
\author[1,3,4]{Roel Verbelen}
\author[1,2,3,4]{Katrien Antonio}
\author[1,3]{Gerda Claeskens}
\author[1,4]{Jonas Crevecoeur}
\affil[1]{Faculty of Economics and Business, KU Leuven, Belgium.}
\affil[2]{Faculty of Economics and Business, University of Amsterdam, The Netherlands.}
\affil[3]{LStat, Leuven Statistics Research Center, KU Leuven, Belgium.}
\affil[4]{LRisk, Leuven Research Center on Insurance and Financial Risk Analysis, KU Leuven, Belgium.}
\date{\today}
\maketitle

\begin{abstract}
A delay between the occurrence and the reporting of events often has practical implications such as for the amount of capital to hold for insurance companies,
or for taking preventive actions in case of infectious diseases.
The accurate estimation of the number of incurred but not (yet) reported events forms an essential part of properly dealing with this phenomenon.
We review the current practice for analysing such data and we present a flexible regression framework to jointly estimate the occurrence and reporting of events.
By linking this setting to an incomplete data problem, estimation is performed via an expectation-maximization algorithm.
The resulting method is elegant, easy to understand and implement, and provides refined insights in the nowcasts.
The proposed methodology is applied to a European general liability portfolio in insurance.
\end{abstract}


\section{Introduction}

This paper reviews and extends the literature on statistical models for problems where individuals (or objects) under study experience two events. The first, also called the initiating or primary, event occurs at time $x$, and the second, the so-called secondary or consequent, event only occurs at a later time $s\geq x$. The presence of the delay $u = s-x$ between the two events leads to statistical challenges, because the currently observed number of primary events is right-censored while observation delays are  right-truncated, see e.g.~the early contributions by \cite{lagakos1988}, \cite{Kalbfleish1989}, \cite{harris1990reporting} and \cite{KalbfleischLawless1991} for an introduction to the field. The term \textit{back-calculation} \citep[][]{Brookmeyer1988,Bacchettietal1993} refers to the reconstruction of the past history of first events that must have occurred to give rise to the observed pattern of second event cases, under the assumption of a known delay distribution. Nowadays, \textit{nowcasting} is often used for estimating the current number of first events using only the available partial information on the reported or registered secondary events \citep[see][for recent examples of nowcasting problems in epidemiology]{Hohle2014,Kassteele2019, Bastos}.

Such delays occur in different ways and in a variety of subject areas. In an insurance setting a claim is only reported some time after its occurrence, because the damage was not immediately noticed or the insured needed some time to file the claim to the insurance company. A proper estimation of these unreported claims is important, since financial regulations force insurance companies to hold sufficient capital reserves to be able to fulfill their future liabilities with respect to such claims. \cite{jewell1989} sketches first contributions to the modelling of these occurred- or incurred-but-not-reported (OBNR or IBNR) claims in an insurance context. In disease modelling at least two examples of these delays are studied in the literature. In the first, $x$ represents the time of diagnosis of a case \citep[or another relevant event, like hospital admission in][]{Donker} and $s$ is the time of reporting of the case to the organization that coordinates the surveillance of the disease. This reporting delay may result from various processes including logistics, or the time to complete a test and to report a confirmed case in a health database. Statistical surveillance systems for the detection of outbreaks of infectious diseases have to properly adjust for reporting delays in order to take timely preventive action \citep[see e.g.][]{Farrington1996,noufaily2015modelling, noufaily2016detection}. In the second example $x$ measures the time of infection (with a virus, for example) and $s$ is the time of onset of disease symptoms. Insight in the distribution of the incubation time $u = s-x$ is important to estimate the current number of infections in a population.
In reliability engineering and quality management, the statistical analysis of warranty data requires taking the time between the failure and its reporting into account in order to predict the number of future warranty claims from all units in service \citep[see e.g.][]{kalbfleisch1991methods, wu2013review}.

The contribution of our paper is threefold. First, Section~\ref{sec:overview} sketches (a selection of recent) contributions on nowcasting that appeared in actuarial, statistical or epidemiological literature. Our literature overview stretches across multiple disciplines and structures these papers along the time scale used in the modelling framework, where we distinguish between models for events in continuous time and models for data aggregated over a coarser discrete time grid. Second, we contribute to the literature by proposing in Section~\ref{sec:EM} a flexible yet practical modelling and estimation framework capable of dealing with any parametric structure for both the occurrence as well as the reporting process. We employ an expectation-maximisation (EM) method \citep{dempster1977maximum} to jointly estimate the occurrence and reporting delay of the events in the presence of covariates, allowing us to acquire the necessary insights in the dynamics of both.
Third, Section~\ref{sec:case_study} presents the results of applying the model in a case study where out-of-time evaluations
are used to asses the predictive performance of the method. Moreover, we benchmark our results against the findings obtained with a selection of modelling strategies from the literature overview. Section~\ref{sec:conclusions} concludes. 
The supplementary document contains some additional results regarding the case study.

\section{Literature overview and notation} \label{sec:overview}

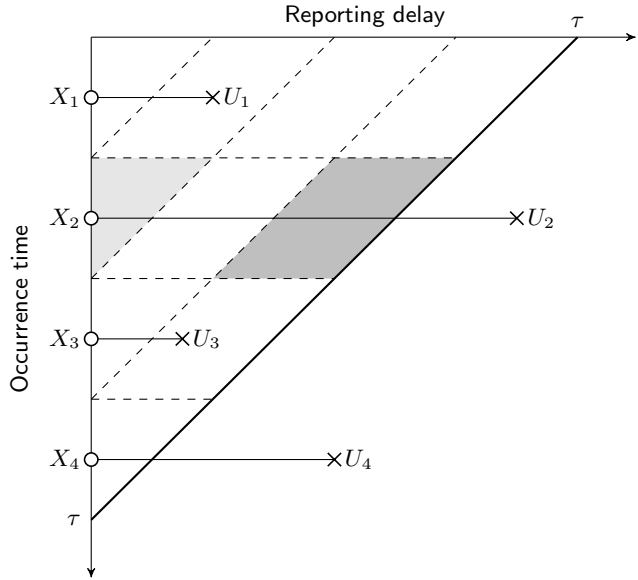
\begin{figure}[htb!]

\begin{center}

\tikzset{cross/.style={cross out, draw,
         minimum size=2*(#1-\pgflinewidth),
         inner sep=0pt, outer sep=0pt}}

\begin{tikzpicture}[scale=0.8, auto, to/.style={->,>=stealth',shorten >=1pt}, every node/.style={font=\fontsize{9pt}{9pt}\selectfont\sffamily, align=center, semithick}]
\path[fill=gray!50] (2, 4) to (4, 4) to (6, 6) to (4, 6);
\path[fill=gray!20] (0, 4) to (0, 6) to (2, 6);
%


\draw[to] (0,8) -- node[above] {Reporting delay} (9,8);
\draw[to] (0,8) -- node[rotate=90, anchor = south, yshift = 0.7cm] {Occurrence time}(0,-1);
\draw[thick] (0,0) -- (8,8);

\draw  (0, 7) node[circle,draw=black, fill=white,inner sep = 0pt, minimum size=5pt] {} -- (2, 7) node[cross,draw=black, fill=white,minimum size=5pt] {};
\draw  (0, 5) node[circle,draw=black, fill=white,inner sep = 0pt, minimum size=5pt] {} -- (7, 5) node[cross,draw=black, fill=white,minimum size=5pt] {};
\draw  (0, 3) node[circle,draw=black, fill=white,inner sep = 0pt, minimum size=5pt] {} -- (1.5, 3) node[cross,draw=black, fill=white,minimum size=5pt] {};
\draw  (0, 1) node[circle,draw=black, fill=white,inner sep = 0pt, minimum size=5pt] {} -- (4, 1) node[cross,draw=black, fill=white,minimum size=5pt] {};

\draw (0, 0) node[left] {$\tau$};
\draw (8, 8) node[above] {$\tau$};

\draw (0, 7) node[left] {$X_1$};
\draw (0, 5) node[left] {$X_2$};
\draw (0, 3) node[left] {$X_3$};
\draw (0, 1) node[left] {$X_4$};

\draw (2, 7) node[right] {$U_1$};
\draw (7, 5) node[right] {$U_2$};
\draw (1.5, 3) node[right] {$U_3$};
\draw (4, 1) node[right] {$U_4$};

\foreach \i in {2, 4, 6}
{
   \draw[dashed] (0, \i) -- (\i, \i);
   \draw[dashed] (0, \i) -- (8-\i, 8);
}

%

\end{tikzpicture}
\end{center}

\caption{
Occurrence and reporting of multiple events over time. Events 2 and 4 are unobserved as these events get reported after the evaluation date $\tau$. The dashed grid indicates the aggregation of the continuous data towards a coarser discrete time scale. The events registered in continuous time in the shaded regions will be aggregated into a single cell or observation in the run-off triangle pictured in Table~\ref{tab:runoff}.}
\label{fig:time_line_rev}
\end{figure}

The event of interest (e.g.~the failure of a product, the occurrence of an insured event or the diagnosis of disease) happens at time $X$, though it is only reported or observed at a later time $S$. The reporting delay is then $U=S-X$.
Figure~\ref{fig:time_line_rev} represents the occurrence of multiple events over time.
When the reporting occurs before $\tau$, the observation is complete (events 1 and 3 in Figure~\ref{fig:time_line_rev}), while reporting after $\tau$ corresponds with a currently unreported case (events 2 and 4 in Figure~\ref{fig:time_line_rev}). At time $\tau$ the analyst has to predict or evaluate the number of events that incurred in the past, but will only be reported in the future. This is challenging because the analyst faces incomplete data, with no information available for the unreported cases. Leaving the continuous time setting, the dashed grid in Figure~\ref{fig:time_line_rev} pictures the aggregation of events when choosing a coarser discrete time scale. The objective is to use the observed, reported events in the upper triangle $\boldsymbol{\Delta}^{\r} = \{(X,U)\ |\ X \leq \tau,\ U \leq \tau - x\}$ to predict unreported events in the lower triangle $\boldsymbol{\Delta}^{\IBNR} = \{(X,U)\ |\ X \leq \tau,\ U > \tau - X\}$.

\subsection{Continuous time models}\label{sec:cont_time}

When data are collected in (almost) real-time, predicting the number of unreported events reduces to estimating the two-dimensional density $f_{X, U}$ for the occurrence and reporting of events in $[0, \tau] \times [0, \tau]$. Hereto, only observations observed on the upper triangle in Figure~\ref{fig:time_line_rev} are available. \cite{Martinez2013} and more recently \cite{Hiabu2020} estimate this density with a two-dimensional kernel density estimator with support on the upper triangle. By assuming a multiplicative kernel, the local linear density estimate can be extrapolated to the lower triangle which then leads to a forecast for the occurred but not yet reported events. In many applications, we are interested in the marginal densities, i.e.~the occurrence intensity of events on the one hand and the reporting delay distribution on the other hand. This interest in the marginal densities is commonly translated into the following assumptions
\begin{enumerate}[label=(A{{\arabic*}})]
	\item \label{A1_cont} The event counting process $N(x)=\sum_{i\ge 1}\mathcal{I}\{X_i\le x\}$ for $x\ge0$ follows an inhomogeneous Poisson point process with an intensity $\lambda(x;\boldsymbol{\alpha})$, which depends on some parameter vector $\boldsymbol{\alpha}$.
	\item \label{A2_cont} Conditional on the occurrence time $X$, the reporting delay $U$ follows a positive continuous distribution with density $f_{U|X}(\cdot; \boldsymbol{\theta})$ and parameter vector $\boldsymbol{\theta}$. The reporting delay $U$ is independent of the event counting process $N(x)$.
\end{enumerate}

\noindent Let $\boldsymbol{\Delta}^{\r}=\{(x_i, u_i),\ i=1,\ldots,n|x_i \leq \tau,\ u_i \leq \tau-x_i\}$ denote the $n$ observed events in the upper triangle. The associated log-likelihood of these reported events becomes
\begin{eqnarray}\label{eq:cont_lik}
\label{eq:full_LL_cont}
\log \mathcal{L}(\boldsymbol{\alpha}, \boldsymbol{\theta}|\boldsymbol{\Delta}^{\r}) = \sum_{(x,u)\in \boldsymbol{\Delta}^{\r}} \log{\left\{\lambda(x;\boldsymbol{\alpha}) \cdot F_{U|X}(\tau - x|x;\boldsymbol{\theta})\right\}} \\
 - \int_{0}^\tau \lambda(v;\boldsymbol{\alpha}) \cdot F_{U|X}(\tau - v|v;\boldsymbol{\theta})dv \nonumber
+ \sum_{(x,u)\in \boldsymbol{\Delta}^{\r}} \log{ \left( \frac{f_{U|X}(u|x;\boldsymbol{\theta})}{F_{U|X}(\tau - x|x;\boldsymbol{\theta})} \right) }.
\end{eqnarray}

\noindent Optimization of this likelihood is complicated by the joint presence of $\lambda(\cdot;\boldsymbol{\alpha})$ from the occurrence process and $F_{U|X}(\cdot|\cdot;\boldsymbol{\theta})$ from the reporting process, prohibiting a split of the likelihood. The likelihood in \eqref{eq:full_LL_cont} already appeared in \cite{Kalbfleish1989}, under the assumption of independence between the time of the initiating event and the duration of the delay. Two strategies for optimizing this likelihood commonly appear in the literature.
First, a strand of research directly optimizes this likelihood, despite its complexity. For example, \cite{Kalbfleish1989} use a parameterized Poisson process for infections and some parametric models for incubation time, allowing the joint estimation of the occurrence and reporting processes in continuous time. In actuarial science \cite{Haastrup} present a Bayesian and \cite{Wahl2019} a frequentist approach with the likelihood in \eqref{eq:full_LL_cont} as focus. The latter contributions build upon the fundamental ideas for the IBNR problem in actuarial science proposed by \cite{jewell1989}, \cite{Norberg1993} and \cite{Norberg1999}.

A second strategy opts for simplicity by proposing a (heuristic) two-stage approach.
For example, \cite{AntonioPlat2014} fit a parametric distribution to the observed reporting delays in a first step and then plug in the estimated reporting delay distribution when estimating the parameters in the thinned Poisson process for the occurrence of reported events. The log-likelihood of the reporting process becomes
$$\log \mathcal{L}(\boldsymbol{\theta}|\boldsymbol{\Delta}^{\r}) = \sum_{(x,u)\in \boldsymbol{\Delta}^{\r}} \log{ \left( \frac{f_{U|X}(u|x;\boldsymbol{\theta})}{F_{U|X}(\tau - x|x;\boldsymbol{\theta})} \right) }.$$
Given the occurrence time $X=x$ of an observed event, its reporting delay is a right-truncated variable $U$ with truncation point $\tau - x$. \cite{lagakos1988}, \cite{KalbfleischLawless1991} and \cite{Lawless1994} (among others) focus on estimating the reporting delay under right-truncation by inverting the direction of time, effectively transforming the right- into left-truncated data. Standard statistical methods for left-truncated data, such as the Cox proportional hazard model \citep{Cox1972}, can then be used to model the reporting of events. \cite{badescu2016marked, badescu2016estimation} and \cite{avanzi2016micro} follow a strategy similar to \cite{AntonioPlat2014}, but model the event occurrence process as a marked Cox process to allow for overdispersion and serial dependency. Along this strand \cite{verrall2016understanding} decouple the full likelihood in \eqref{eq:full_LL_cont} by considering a plug-in estimate for the weekly periodic occurrence pattern of insurance claims, followed by estimating parametric distributions for the reporting delay.

\subsection{Discrete time models}\label{sec:discrete_time_models}

\subsubsection{Aggregating events towards a coarser time scale.}

Events are usually not registered in continuous, but in discrete time periods. Such discrete event counts result by aggregating the real time events towards a coarser time scale. Figure~\ref{fig:time_line_rev} sketches this aggregation from continuous to discrete time data. The day is the natural (discrete) time unit in many administrative systems. Daily data are often aggregated to weekly, monthly, quarterly or yearly event data, even though many granular data insights may get lost in this aggregation process. For example, \citet{BeckerCui1997} consider discrete time units in quarterly periods and model the delay in entering AIDS diagnoses in a surveillance system. Insurance studies traditionally use quarterly or yearly claim counts \citep{WuthrichMerz2008}.
In our notation and in the case study presented in Section~\ref{sec:case_study} we consider event counts on a daily level, even though our notation is generic and may refer to any other time unit of interest (e.g., weeks, months or years).
The number of events which occurred on day $t$ is denoted by $N_t$, where the integer $t$ indicates the occurrence date and ranges from 1 to ${\tau}$. The number of these events which have been reported after precisely $d$ days are denoted as $N_{td}$ such that
$ N_{t} = \sum_{d=0}^{\infty} N_{td}.$ Only those events that are reported before or at the evaluation date ${\tau}$ are observed, causing $N_{t}$ to be right-censored and the reporting delay distribution of the observed events occurring on day $t$ to be right-truncated at $\tau - t$. Throughout this paper we denote the reported, and therefore observed, number of events which occurred on day $t$ by
$$   N_{t}^{\r} = \sum_{d=0}^{\tau - t} N_{td}.  $$
We denote the number of events that have happened on day $t$ but are not yet reported by
\[  N_{t}^{\IBNR} = \sum_{d=\tau - t + 1}^{\infty} N_{td}  \, ,  \]
and the total number of such unreported events over all occurrence days by
\[  N^{\IBNR} = \sum_{t = 1}^{\tau} N_{t}^{\IBNR} = \sum_{t = 1}^{\tau} \sum_{d=\tau - t + 1}^{\infty} N_{td}.
\]
In discrete time nowcasting the overall objective is to use the observed events $\bs{N}^{\r} = \{  N_{td} \mid 1 \leq t \leq \tau, d \geq 0, t + d \leq \tau \}$ to predict the events that are not yet reported $\bs{N}^{\IBNR} = \{  N_{td} \mid 1 \leq t \leq \tau, d > 0, t + d > \tau \}$.

A convenient way to represent these event counts $N_{td}$ is with a two-way contingency table where the rows correspond to occurrence time $t$ and the columns indicate the reporting delay $d$. Since we can only observe what has been reported, an incomplete two-way contingency table results where the upper left-hand \emph{triangle} is observed and the lower right-hand \emph{triangle} is empty and has to be predicted. Table~\ref{tab:runoff} depicts the structure of such a triangle with reported events, often called a run-off triangle in actuarial literature or a reporting triangle in epidemiology. The shaded regions in Figure~\ref{fig:time_line_rev} and Table~\ref{tab:runoff} visualize the aggregation of events registered in continuous time into a single cell or observation in this triangle. The dimension of the triangle depends on the granularity of the discretization. Smaller bin sizes correspond to larger triangles, which usually result in more complex models to capture the heterogeneity present in the data.

\begin{center}
\begin{table}[ht]
\caption{\label{tab:runoff}
Triangular display with aggregated event counts. Only the event counts in the upper triangle are observed, whereas the occurred but not reported counts in the lower triangle have to be predicted. The coloured cells in the triangle represent the aggregation of events registered in continuous time in Figure~\ref{fig:time_line_rev} towards a coarser discrete time scale.}
\renewcommand{\arraystretch}{1.3}
\begin{center}
\begin{tabular}{|c| p{0.9cm} p{0.9cm} p{0.9cm} p{0.9cm} p{0.9cm} | }
\hline
Occurrence & \multicolumn{5}{c|}{Reporting delay} \\
period & 0 & $\cdots$ &  $\tau-t$  & $\cdots$\ & $\tau-1$ \\
\hline
1 & $N_{10}$  &  $\cdots$ &  $N_{1, \tau-t}$  &  $\cdots$ &  $N_{1,\tau-1}$ \\
\cline{6-6}
\vdots & &  &  & \multicolumn{1}{c|}{} &  \\
\cline{5-5}
$t$ & \multicolumn{1}{c}{\cellcolor{gray!20}$N_{t0}$}  &  $\cdots$ &  \multicolumn{1}{c|}{\cellcolor{gray!50}$N_{t, \tau-t}$} &    &   \\
\cline{4-4}
$\vdots$ &   & \multicolumn{1}{c|}{} &  & &    \\
\cline{3-3}
$\tau$ & \multicolumn{1}{c|}{$N_{\tau 0}$} &  & & &  \\
\cline{2-2}
\hline
\end{tabular}
\end{center}
\end{table}
\end{center}

\subsubsection{A regression structure for the reporting intensity.}\label{sec:discrete_regr}

A first modelling strategy directly specifies a count regression model for the observed counts $N_{td}$ in $\bs{N}^{\r}$. Most often these counts are assumed to be independent and Poisson distributed, i.e.,~
$ N_{td} \sim \textrm{POI}( \lambda_{td}),
$
or, in case of overdispersion, a negative binomial distribution is used, i.e.,~
$
N_{td} \sim \textrm{NB}( \lambda_{td},\phi),
$
where $\lambda_{td}$ is the reporting intensity. In epidemiology, \cite{Kassteele2019} model the two-dimensional surface in Table~\ref{tab:runoff} using bivariate P-splines with a smooth reporting intensity $\lambda_{td}$ and day-of-the-week effects expressed as deviations from it. This intensity is modelled via a tensor product B-spline basis with pairwise products $B_i(t)B_j(d)$, where the basis functions $B_i(t)$ capture the effect of occurrence time and the $B_j(d)$ model the reporting delay. For the day-of-the-week effects dummy variables are included in the predictor, taking the value 1 if a certain combination of $t$ and $d$ corresponds to a specific weekday, and 0 otherwise. The smooth intensity and day of the week effects are estimated simultaneously using a penalized negative binomial regression model and the iterative reweighted least squares algorithm. \cite{Bastos} correct for reporting delays in disease surveillance data with a negative binomial regression model for the $N_{td}$ that takes spatiotemporal variation into account. In their Bayesian framework the occurred-but-not-reported cases are estimated from the resulting posterior predictive distribution. A similar Bayesian modelling strategy is proposed in \cite{McGough2020}, which is then applied in \cite{Greene2021} to nowcast COVID-19 cases in New York City.

In actuarial science the so-called chain ladder method is probably the most widely used technique to predict numbers of unreported claims. In this method the incremental event counts $N_{td}$ are independently Poisson distributed \citep{hachemeister1975ibnr, renshaw1998stochastic} with $\E [N_{td}] = \lambda_{td} = \lambda_t \cdot p_{d}$ for all $t = 1, \ldots, \tau$ and $d = 0, \ldots, \tau-1$, where $\lambda_1,\ldots,\lambda_{\tau}>0$ and $p_0, \ldots, p_{\tau-1}>0$ with sum constraint $p_0 + \ldots + p_{\tau-1} = 1$. The latter assumes all claims to be reported by the end of the delay period $\tau -1$. Beyond actuarial science, a similar modelling strategy was also proposed in (for instance) the disease modelling examples in \cite{Kalbfleish1989} and \cite{BeckerCui1997}.
The log-likelihood corresponding to the Poisson model formulation 
of the chain ladder method \citep[also referred to as the Poisson log-linear model in][]{Sellero1996} is
\begin{eqnarray*}
\log \mathcal{L}(\bs{\Psi} ; \bs{N}^{\r}) & = & \sum_{t = 1}^{\tau} \sum_{d=0}^{\tau - t}\left\{  -\lambda_t \cdot p_{d} + N_{td} \log(\lambda_t \cdot p_{d})\right\}+c,
\end{eqnarray*}
where we denote $\bs{\Psi} = \{  \lambda_1, \ldots, \lambda_{\tau} , p_{0}, \ldots, p_{\tau-1}  \}$ and $c$ is a constant not depending on the model parameters. Maximizing the log-likelihood requires solving the following set of equations for $\bs{\Psi}$,
\begin{eqnarray}
\sum_{d = 0}^{\tau - t}  {\lambda}_t  \cdot {p}_{d}   & = & \sum_{d = 0}^{\tau - t}  N_{td} \, ,	\qquad		t = 1, \ldots, \tau \, , \label{eq:CL_i}  \\
\sum_{t = 1}^{\tau - d}  {\lambda}_t   \cdot  {p}_{d}  & = & \sum_{t = 1}^{\tau - d}  N_{td} \,	, \qquad		d = 0, \ldots, \tau-1 \, , \label{eq:CL_j}
\end{eqnarray}
subject to all elements of $\bs{\Psi}$ being positive and $\sum_{d=0}^{\tau-1} {p}_{d} = 1$. The Poisson maximum likelihood conditions equate the sums of the claim counts in each row and column of the observed upper triangle $\bs{N}^{\r}$ to their expected value counterparts.
\cite{mack1991simple, mack1993distribution} points out that this set of equations can be solved recursively due to the triangular structure.
A more standard approach to estimate the parameters is to formulate the model as a generalized linear model and use a numerical optimizer \citep[see][]{Taylor2000, EnglandVerrall2002, WuthrichMerz2008}.

\subsubsection{Modelling the occurrence and reporting processes.}\label{sec:flex_models_discrete_time}

An alternative approach explicitly models the underlying occurrence and reporting processes, in line with the strategy outlined in the continuous time setting. The framework is then hierarchical where the $N_t$ follow an occurrence model and the $N_{td}|N_t$ are multinomially distributed. In discrete time the assumptions \ref{A1_cont}-\ref{A2_cont} become
\begin{enumerate}[label=(A{{\arabic*}}')]
	\item \label{A1} The daily total event counts $N_{t}$ for $t = 1, \ldots, \tau$ are independently Poisson distributed with intensity ${\lambda}_{t} = \exp(\bs{x}_{t}' \bs{\alpha})$, where $\bs{x}_{t}$ is the vector of covariate information corresponding to occurrence day $t$ and $\bs{\alpha}$ is a parameter vector.
	\item \label{A2} Conditional on $N_{t}$, the event counts $N_{td}$ for $d = 0, 1, 2, \ldots$, are multinomially distributed with probabilities $p_{td}=p_{td}(\bs{\theta},\bs{x}_{td})$. These reporting probabilities do not depend on the number of events that occurred on day~$t$,
sum to one and are modeled with parameter vector $\bs{\theta}$ and covariate information $\bs{x}_{td}$.
\end{enumerate}
The discrete occurrence intensities and reporting probabilities can be retrieved from the continuous time assumptions in $(A1), (A2)$, when we assume $\lambda_{t}$ to be piecewise constant between integer time points and properly integrate the reporting density $f_{U|X}(\cdot|\cdot;\boldsymbol{\theta})$. That is,
\begin{align*}
	\lambda(x) &= \lambda_t \quad\ \text{for}\ x \in [t-1,t), \\
	p_{td} &= \begin{cases}
	\int_{t-1}^{t} \left( \int_{0}^{t - x} f_{U|X}(u|x;\boldsymbol{\theta}) \, du \right) \, dx & \text{for}\ d = 0 \\
	\int_{t-1}^{t} \left( \int_{d - 1 + t - x}^{d + t - x} f_{U|X}(u|x;\boldsymbol{\theta}) \, du \right) \, dx & \text{for}\ d > 0. \\
	\end{cases}
\end{align*}
The split in cases with reporting delay zero and a strictly positive delay follows from the different shapes (i.e., the triangle in light gray versus the parallelogram in dark gray) when aggregating continuous data in Figure~\ref{fig:time_line_rev}. Note that the chain-ladder method discussed in Section~\ref{sec:discrete_regr} fits in the framework outlined by \ref{A1}-\ref{A2} if we impose the sum constraint $p_0 + \ldots + p_{\tau-1} = 1$. The latter assumes all claims to be reported by the end of delay period $\tau -1$, an assumption typically made in an insurance context. The chain-ladder method assumes a stationary reporting process since the probabilities $p_{d}$ do not depend on the occurrence period $t$.

We bundle the parameters to be estimated in $\bs{\Theta} =  (\bs{\alpha}, \bs{\theta})$. Based on the model assumptions \ref{A1}-\ref{A2} and the thinning property of Poisson random variables, the daily event counts $N_{td}$ are independently Poisson distributed with intensities $\lambda_t(\bs{x}_{t}' \bs{\alpha})\cdot p_{td}(\bs{\theta},\bs{x}_{td})$ for $t = 1, \ldots, \tau$ and $d = 0, 1, 2, \ldots$. Below we (often) use the more compact notation $\lambda_t(\bs{x}_{t}' \bs{\alpha})\cdot p_{td}(\bs{\theta},\bs{x}_{td}) =\lambda_t \cdot p_{td}$ to make our writing more concise.
In particular, the observed event count $N_t^{\r}$ on day $t$ is Poisson distributed with intensity $\lambda_t \cdot p_{t}^{\r}$ where $p_{t}^{\r} =  \sum_{d=0}^{\tau - t} p_{td}$ and the unreported event count $N_t^{\IBNR}$ is Poisson distributed with intensity $\lambda_t \cdot p_{t}^{\IBNR}$ where $ p_{t}^{\IBNR} =  1 - p_{t}^{\r}$. Conditional on $N_t^{\r}$, the observed daily event counts $\{ N_{td} \mid d = 0, 1, \ldots, \tau-t \}$ are multinomially distributed with parameters $N_t^{\r}$ and $\{ p_{td} / p_{t}^{\r} \mid d = 0, 1, \ldots, \tau-t \}$, since we have to account for the right-truncation of the reporting delay. The likelihood of the observed data can then be written as the product of a Poisson likelihood and a multinomial likelihood,
\begin{equation}
\mathcal{L}(\bs{\Theta} ; \bs{N}^{\r}) = \prod_{t = 1}^{\tau} \frac{\exp(-\lambda_t p_{t}^{\r}) \left(\lambda_t p_{t}^{\r} \right)^{N_t^{\r}} } { N_t^{\r}! } \frac{ N_t^{\r}! }  { \prod_{d=0}^{\tau - t}  N_{td} ! } \prod_{d=0}^{\tau - t} \left( \frac{p_{td}}{p_{t}^{\r}} \right) ^{N_{td}} \, .
\label{eq:observed_likelihood}
\end{equation}
Equivalently, the likelihood can also be constructed by treating the daily event counts as right-censored, since the number of unreported events is unknown,
\begin{eqnarray*}
\mathcal{L}(\bs{\Theta} ; \bs{N}^{\r})  & = & \prod_{t = 1}^{\tau} \sum_{n = N_t^{\r}}^{\infty} \frac{\exp(-\lambda_t) \lambda_t ^{n}}{n!} \frac{n!}{(n - N_t^{\r})!\prod_{d=0}^{\tau - t}  N_{td}!} \prod_{d=0}^{\tau - t} \left( p_{td} \right) ^{N_{td}} \left( p_{t}^{\IBNR} \right)^ {n - N_t^{\r}}  \, .
\end{eqnarray*}
Indeed, this expression reduces to \eqref{eq:observed_likelihood} by rewriting the sum over $n$ using the Taylor expansion for the exponential function. The corresponding log-likelihood equals
\begin{equation}
\log \mathcal{L}(\bs{\Theta} ; \bs{N}^{\r})  =  \sum_{t = 1}^{\tau}\Big[-\lambda_t p_{t}^{\r} +  N_t^{\r} \log(\lambda_t) + \sum_{d=0}^{\tau - t}\{ N_{td} \log(p_{td})  -  \log(N_{td} !)\}  \Big].
\label{eq:observed_log_likelihood}
\end{equation}
Similar to the continuous time setting, the log-likelihood in \eqref{eq:observed_log_likelihood} contains terms which depend on the parameters of both the Poisson model for event occurrences (appearing in $\lambda_t$) and the reporting delay distribution (appearing in $p_{td}$ and $p_{t}^{r}$). This complicates direct maximum likelihood estimation as it prevents separate optimization with respect to each of these parameter components. In line with the continuous time literature review in Section~\ref{sec:cont_time}, two strategies for optimizing this likelihood are generally applied. A first strand puts focus on direct optimization of this likelihood using a standard numerical method such as Newton-Raphson. This is feasible, but we cannot rely on standard statistical modelling routines and we need to derive (analytically or numerically) the gradient and Hessian of the log-likelihood in \eqref{eq:observed_log_likelihood}. \cite{Hohle2014} propose a Bayesian approach for the joint modelling of the occurrence and delay process for discrete data collected in epidemiology. \cite{Gunther2021} then apply this Bayesian model to construct nowcast estimates for the COVID-19 pandemic in Bavaria. In a second strand, a (heuristic) two-stage approach keeps the likelihood tractable by putting focus on the modelling of the reporting process, while using a nonparametric estimator for the occurrence process. Examples are \cite{Lawless1994} on random temporal fluctuations in reporting delays and \cite{crevecoeur2019} on the reporting of insurance claims in the presence of holiday effects.

\subsection{The EM algorithm to tackle the missing data problem}

Our literature overview for both continuous as well as discrete time models puts focus on either the direct optimization of the likelihood or the use of heuristic two-stage procedures. In the absence of right-truncation the complete likelihood analogue of \eqref{eq:cont_lik} (continuous) or \eqref{eq:observed_log_likelihood} (discrete) would split into an occurrence and a reporting likelihood and become tractable. This immediately suggests the use of the expectation-maximisation (EM) algorithm \citep{dempster1977maximum} to handle the missing data problem, as is recognized by a series of contributions on nowcasting. \cite{Brookmeyer1988} estimate a piecewise constant infection density while assuming a Weibull distribution for the incubation time, independent of the time of infection. With an iterative procedure similar to EM, \cite{harris1990reporting} fits both categorical as well as continuous-time models for the incidence of infections and their incubation time. \cite{Pagano1994} use the EM algorithm but only focus on the estimation of the reporting delay distribution, while \citet{Bacchettietal1993} and \cite{Bellocco2000} assume a known (incubation) distribution for the delays. These papers propose rather simple structures for the occurrence and reporting processes, e.g.~a stationary reporting delay distribution, or categorical models with separate parameters for each $t$ and $d$. In the next section we present a flexible yet practical estimation framework capable of dealing with any parametric structure for the occurrence and reporting processes. Because data are collected over discrete
time units in administrative systems, we propose our general model for the intensities $\lambda_t(\bs{x}_{t}' \bs{\alpha})$ and the reporting delay probabilities $p_{td}(\bs{\theta},\bs{x}_{td})$ in \ref{A1}-\ref{A2}. This framework not only allows for an easy calibration of each of the models from Section~\ref{sec:flex_models_discrete_time}, but also facilitates the estimation of new occurrence and reporting process specifications which were previously considered too difficult for joint optimization. Aiming for flexibility, we incorporate covariates in both the occurrence process of events and the reporting delay distribution. We capture evolutions over time or seasonal trends, by including fluctuations in both the event counts and their reporting delays by month, day of the month or day of the week of the occurrence date. Additionally, one can also model relationships with external covariates which might influence the events such as economic circumstances, business cycles and weather conditions. Day-specific particularities in the reporting delay (such as holiday effects) can be modeled using designated day probabilities. The case-study developed in Section~\ref{sec:case_study} convincingly illustrates this, by letting the data set assist us in choosing an appropriate occurrence and reporting delay structure.

\section{An EM algorithm for the joint estimation of the occurrence and reporting delay of events}
\label{sec:EM}

\subsection{Parameter estimation using an EM algorithm} \label{sec:EM_reservingModel}

Starting from the log-likelihood in \eqref{eq:observed_log_likelihood} we choose to treat the truncation as a missing data problem and employ an EM algorithm 
to simplify maximum likelihood parameter estimation.
Consider the complete version of the data $\bs{N} = \bs{N}^{\r} \cup \bs{N}^{\IBNR} = \{  N_{td} \mid 1 \leq t \leq \tau,  d \geq 0\}$ which augments the observed daily event counts from the upper part of the triangular display  in Table \ref{tab:runoff} with the unknown values of the future event counts in the lower triangle. Given the complete data $\bs{N}$, we construct the complete log-likelihood function
\begin{eqnarray}
\qquad \log \mathcal{L}_c(\bs{\Theta} ; \bs{N}) =  \sum_{t = 1}^{\tau}\Big[ -\lambda_t +  N_t \log(\lambda_t) +  \sum_{d=0}^{\infty} \{N_{td}\log(p_{td}) - \log(N_{td} !)\} \Big],
\label{eq:complete_log_likelihood}
\end{eqnarray}
which allows for a separate estimation of the parameters $\boldsymbol{\alpha}$ of the event occurrence model (appearing in $\lambda_t(\bs{x}_{t}' \bs{\alpha})$) and $\boldsymbol{\theta}$ of the reporting delay distribution (appearing in $p_{td}(\bs{\theta},\bs{x}_{td})$).

From a numerical point of view, we propose to deal with the infinite sums over the reporting delay $d$ in $\eqref{eq:complete_log_likelihood}$ by introducing
$$N_{t, \tau+} = \sum_{d \geq \tau} N_{t,d}$$
and
$$N^{\texttt{nr}} = \{N_{t,d} \mid 1 \leq t \leq \tau, t+d > \tau, d < \tau \} \cup \{N_{t, \tau+} \mid 1 \leq t \leq \tau\}.$$
This implies grouping all events reported with a delay of at least $\tau$ days in a remainder category. The latter are the events reported beyond the right boundary of the reporting triangle in Table~\ref{tab:runoff}. The complete log-likelihood still factorizes into occurrence and reporting contributions and becomes
\begin{eqnarray}
\qquad \log \mathcal{L}_c(\bs{\Theta} ; \bs{N}) &=& \sum_{t = 1}^{\tau}\Big[ -\lambda_t +  N_t \log(\lambda_t) +  \sum_{d=0}^{\tau-1} \{N_{td}\log(p_{td}) - \log(N_{td} !)\}  \nonumber \\
&& + N_{t,\tau+} \cdot \log\Big( \sum_{d \geq \tau} p_{td} \Big) - \log(N_{t,\tau+}!) \Big]. \label{eq:complete_log_likelihood_adj}
\end{eqnarray}
The EM algorithm exploits the simpler form of $\mathcal{L}_c(\bs{\Theta} ; \bs{N})$ by iterating between computing expectations in an E-step and maximization in an M-step.
Applied to our setting, the numbers of unreported events are replaced by their expected values in the E-step and the log-likelihood of the augmented data is maximized in the M-step. While numerical optimization is required in the M-step, the parameters of the event occurrence model can be estimated separately from the reporting delay parameters and standard software routines can be utilized in the absence of truncation. We discuss the steps of the EM algorithm in more detail for the $k$th iteration.

\paragraph{E-step.} We take the conditional expectation of the complete log-likelihood \eqref{eq:complete_log_likelihood_adj} given the observed data $\bs{N}^{\r}$ and using the current estimate $\bs{\Theta}^{(k-1)}$ of the parameter vector  $\bs{\Theta}$:
\begin{align}   Q(\bs{\Theta}; \bs{\Theta}^{(k-1)})  = \E_{\bs{\Theta}^{(k-1)}} \left[ \log \mathcal{L}_c(\bs{\Theta} ; \bs{N}) \mid  \bs{N}^{\r}\right].
\label{eq:Q_reserving}
\end{align}
This requires to compute the expected values of the future event counts
\begin{equation}
 N_{td}^{(k)} = \E_{\bs{\Theta}^{(k-1)}} \left[ N_{td} \mid   \bs{N}^{\r} \right] =
\begin{cases}
N_{td} & \mbox{if } d \leq \tau - t \\
\lambda_t^{(k-1)} p_{td}^{(k-1)} & \mbox{otherwise,}  \\
\end{cases}
\label{eq:Nk}
\end{equation}
for $t = 1, \ldots, \tau$ and the total daily event counts are
$$
N_{t}^{(k)} = \sum_{d=0}^{\tau-t} N_{td}  + \sum_{d=\tau - t + 1}^{\tau-1} N_{td}^{(k)} + N_{t, \tau+}^{(k)}.
$$
The terms in \eqref{eq:Q_reserving} containing $\E_{\bs{\Theta}^{(k-1)}} \left[ \log (N_{td}!) \mid  \bs{N}^{\r}  \right]$ depend on the observed event counts and the parameter values in iteration $k-1$ and these will therefore be treated as constants in the M-step where the maximization is with respect to the parameter vector $\bs{\Theta}$.

\paragraph{M-step.} We maximize the expected value \eqref{eq:Q_reserving} of the complete data log-likelihood obtained in the E-step with respect to the parameter vector $\bs{\Theta}$.
In order to optimize \eqref{eq:Q_reserving} with respect to $\bs{\alpha}$ as defined in model assumption \ref{A1}, we have to maximize the terms related to the event occurrence model,
\begin{equation}
- \sum_{t = 1}^{\tau} \lambda_t  +  \sum_{t = 1}^{\tau} N_t^{(k)} \log(\lambda_t )  = - \sum_{t = 1}^{\tau} \exp(\bs{x}_t' \bs{\alpha}) +  \sum_{t = 1}^{\tau} N_t^{(k)} \bs{x}_t' \bs{\alpha} \, ,
\label{eq:Q_alpha}
\end{equation}
which is a weighted Poisson log-likelihood, possibly including an offset term (to include an exposure-to-risk measure, as illustrated in the case study covered in Section~\ref{sec:case_study}). The parameter values optimizing \eqref{eq:Q_alpha} are denoted by $\bs{\alpha}^{(k)}$.
Updating the estimates for the parameters $\bs{\theta}$ of the reporting delay distribution requires the maximization of
\begin{eqnarray}
\sum_{t = 1}^{\tau} \sum_{d=0}^{\tau-1} N_{td}^{(k)} \log\{p_{td}(\bs{\theta},\bs{x}_{td})\} + \sum_{t=1}^{\tau} N_{t,\tau+}^{(k)} \cdot \log\{\sum_{d \geq \tau} p_{td}(\bs{\theta},\bs{x}_{td}) \}. \label{eq:Q_theta}
\end{eqnarray}
We now have to optimize a
right-censored likelihood with right-censoring at delay $\tau$. Depending  on the problem at hand, the likelihood can be simplified by omitting the censoring term when reporting events with a delay of more than $\tau-1$ days is highly uncommon, see our discussion in the case study of Section~\ref{sec:case_study}.

\paragraph{Initial step.} The first E-step of the EM algorithm with $k=1$ requires a starting value $\bs{\Theta}^{(0)}$ for the parameter set. Our strategy is to first apply the chain ladder method (see Section~\ref{sec:discrete_time_models}) on the daily event counts to obtain initial predictions $N_{td}^{(0)}$ of the future event counts. Then, we initialize $\bs{\Theta}$ by applying an M-step based on these initial event count estimates. More specifically, we use the \cite{WuthrichMerz2015} formulation of the chain ladder method in terms of the chain ladder development factors, allowing for a fast and practical implementation. Therefore, we define the cumulative event counts as
\[   C_{td} = \sum_{j=0}^{d} N_{tj} \quad \mbox{for } t = 1, \ldots, \tau, \mbox{and }d = 0, \ldots, \tau-t \, , \]
and estimate the development factors of the chain ladder method as
\begin{equation}
\widehat{f}_{d} = \frac{\sum_{t = 1}^{\tau - d} C_{td} }{\sum_{t = 1}^{\tau - d} C_{t, d-1} } \quad \mbox{for } d = 1, \ldots, \tau-1 \, .
\label{eq:chainladder_factors}
\end{equation}
The chain ladder method applies these development factors to the latest cumulative event count in each row to produce forecasts of future cumulative event counts:
\begin{eqnarray*}  \widehat{C}_{t, d} =  C_{t, \tau-t} \widehat{f}_{\tau-t + 1} \ldots  \widehat{f}_{d} \quad \mbox{for }  t = 2, \ldots, \tau, \mbox{and } d = \tau - t +1, \ldots, \tau-1 \, .
\label{eq:chainladder_forecast}
\end{eqnarray*}
We use these chain ladder estimates for the cumulative event counts to initialize the expected incremental event counts as
\[
 N_{td}^{(0)} =
\begin{cases}
N_{td} & \mbox{if } d \leq \tau - t \\
\widehat{C}_{t, \tau-t+1} - C_{t, \tau-t} & \mbox{if } d = \tau-t+1 \\
\widehat{C}_{td} - \widehat{C}_{t, d-1}  & \mbox{otherwise,}  \\
\end{cases}
\]
for $t = 1, \ldots, \tau$ and apply an M-step, as outlined above with $k=0$, to find decent starting values $\bs{\Theta}^{(0)}$. For delays beyond the maximum observed delay (i.e.~$d > \tau-1$) we put the initial values $N_{td}^{(0)}$ equal to zero.

\paragraph{Convergence.}
The log-likelihood \eqref{eq:observed_log_likelihood} increases with each EM iteration \citep{dempster1977maximum}.
Given proper starting values, the sequence $\bs{\Theta}^{(k)}$ converges to the maximum likelihood estimator (MLE) of $\bs{\Theta}$ corresponding to the incomplete data log-likelihood $\log \mathcal{L}(\bs{\Theta} ; \bs{N}^{\r})$ in \eqref{eq:observed_log_likelihood}.
The stopping criterion we apply is based on the relative change in the log-likelihood. Namely, we iterate until the absolute value of
$ \{\log \mathcal{L}(\bs{\Theta} ^{(k)} ; \bs{N}^{\r})   -  \log \mathcal{L}(\bs{\Theta}^{(k-1)} ; \bs{N}^{\r}) \}/ \{ 0.1 +  \log \mathcal{L}(\bs{\Theta} ^{(k)} ; \bs{N}^{\r})  \} $
becomes smaller than $10^{-8}$.
The parameter vector estimate upon convergence is denoted by
$ \widehat{\bs{\Theta}}$.

\subsection{Asymptotic variance-covariance matrix} \label{sec:asymp_cov}

It is expected that when the number of parameters is finite, asymptotic normality of the estimators holds under quite general conditions, though they depend on the specific problem settings. However, if the number of parameters increases with $\tau$, such as in a chain ladder model, the asymptotic properties of the maximum likelihood estimators are difficult to obtain and require a more detailed and rigorous study.


While an EM algorithm by itself does not output an estimated covariance matrix of the parameter estimators, with some additional effort, such matrix may be constructed.
\cite{Louis1982} showed how the observed information matrix can be expressed in terms of the gradient and second derivative of the complete data log-likelihood function.
This approach avoids working with the incomplete data log-likelihood function and leads to analytic expressions which can be computed using the model specifications.
The supplemented EM algorithm \citep{MengRubin1991} provides a computational, iterative approach using
the negative second Hessian matrix of the expected complete-data log likelihood.
\citet{Oakes1999} explains how the matrix of second derivatives of the observed data log-likelihood can be obtained using the derivatives of the function $Q$, see \eqref{eq:Q_reserving}, and provides expressions for exponential family models.
For models for which analytical results are too hard to obtain, numerical derivatives could be used instead.

\subsection{Variable selection and model diagnostics}
\label{ssec:AICdiagnostics}

If one works with the log-likelihood from \eqref{eq:observed_log_likelihood}, in principle all known methods for likelihood estimation are applicable, in particular the use of information criteria such as AIC \citep{Akaike73} and BIC \citep{Schwarz78} for variable selection, see \citet[][Chap.~2, 3]{ClaeskensHjort2008} for more explanation about their use.
However, when an EM algorithm is used for estimation, care needs to be taken since at convergence the function $Q$ of \eqref{eq:Q_reserving} is not the maximized log-likelihood.
\citet{CavanaughShumway1998} developed AICcd which allows to directly work with $Q$ though it requires an adjustment to the `penalty' part of the AIC.
Such an adjustment is used by \citet{ClaeskensConsentino2008} for variable selection with missing covariate data, and by \citet{DirickClaeskensBaesens2015} for selection in multiple event mixture cure models.

The complete data AIC value for use with an EM algorithm is defined by
$$ \text{AICcd} = -2 Q(\widehat{\bs{\Theta}};\widehat{\bs{\Theta}})+2\text{trace}\{
{I}_c(\widehat{\bs{\Theta}};\bs{N}^{\r}) {I}^{-1}_o(\widehat{\bs{\Theta}};\bs{N}^{\r})\},
$$
with the square matrices
$$
{I}_c(\widehat{\bs{\Theta}}; \bs{N}^{\r})  =
- \frac{ \partial^2 }{ \partial \bs{\Theta} \partial \bs{\Theta}' } Q(\widehat{\bs{\Theta}};\widehat{\bs{\Theta}});
\quad
{I}_o(\widehat{\bs{\Theta}};\bs{N}^{\r}) =
 - \frac{ \partial^2 }{ \partial \bs{\Theta} \partial \bs{\Theta}' } \log \mathcal{L}(\widehat{\bs{\Theta}};\bs{N}^{\r}),
$$
where $I_o$ estimates the limiting covariance matrix of $\widehat{\bs{\Theta}}$, for which methods such as described in
Section~\ref{sec:asymp_cov} may be used.

For model selection purposes, the AICcd values are computed for each considered model and the model with the smallest value of AICcd gets selected.

The matrix  ${I}_c(\widehat{\bs{\Theta}};\bs{N}^{\r})$  is also the main ingredient for model diagnostic measures after using an EM algorithm, see \citet{ZhuLeeWeiZhou2001} and \citet{BarretoSouzaSimas2017}. To define the generalized Cook's distance to investigate the influence of a single observation, thus of a single $N_{td}$ with $t=1,\ldots,\tau$ and $d=0,\ldots,\tau-1$, we use a one-step approximation as in \citet{ZhuLeeWeiZhou2001} to avoid refitting the model. Define
$\dot Q_{-[t,d]}({\bs{\Theta}};\widehat{\bs{\Theta}})
= E[\partial\log \mathcal{L}_{c,-[t,d]}(\bs{\Theta} ; \bs{N})/(\partial\bs{\Theta})]
$
with observation $t,d$ removed from the summation, which is estimated by replacing ${\bs{\Theta}}$ by $\widehat{\bs{\Theta}}$. The generalized Cook's distance for use with the output of an EM algorithm now reads, for $t=1,\ldots,\tau$ and $d=0,\ldots,\tau-1$,
\begin{eqnarray*}
\text{GD}_{t,d}  = 
\dot Q_{[t,d]}(\widehat{\bs{\Theta}};\widehat{\bs{\Theta}})'
{I}_c^{-1} (\widehat{\bs{\Theta}};\bs{N}^{\r}) \dot Q_{[t,d]}(\widehat{\bs{\Theta}};\widehat{\bs{\Theta}}).
\end{eqnarray*}
Large values should encourage further investigation of those observations.
We refer to the mentioned references for other influence measures that can be used with an EM algorithm.

\subsection{Prediction and out-of-time evaluation tools} \label{sec:prediction}

Using the estimated parameter vector $\widehat{\bs{\Theta}}$, we predict the number of daily unreported events in the lower triangle of Table \ref{tab:runoff}. Point estimators for all $N_{td} \in \bs{N}^{\IBNR}$ can be obtained using the expected values $\widehat{N}_{td} =  
\widehat{\lambda}_t \, \widehat{p}_{td}$. Similarly, the total number of unreported events per day is estimated by $\widehat{N}_{t}^{\IBNR} = \sum_{d=\tau - t + 1}^{\infty} \widehat{N}_{td} = \widehat{\lambda}_t \,  \widehat{p}_{t}^{\IBNR}$ and the total number of unreported events over all occurrence days by $\widehat{N}^{\IBNR} = \sum_{t = 1}^{\tau} \widehat{N}_{t}^{\IBNR}$.
Moreover, under the model assumptions \ref{A1}-\ref{A2}, the future daily number of events $N_{td}$ $(t \leq \tau, t+d > \tau)$ are independently Poisson distributed and we thus have that
$N_{td}  \sim \text{Poisson}(\lambda_t \, p_{td})$, $N_t^{\IBNR}  \sim \text{Poisson}( \lambda_t  \, p_{t}^{\IBNR})$
and $N^{\IBNR} \sim \text{Poisson} \left(\sum_{t = 1}^{\tau} \lambda_t  \, p_{t}^{\IBNR} \right)$.
This allows us to construct prediction intervals and to make probabilistic statements concerning the number of unreported events after replacing the intensities by their maximum likelihood estimates. With a model defined on a daily level (by means of example) these unreported events can be divided into daily nowcasts by occurrence date or by reporting date, as we demonstrate in the case study in Section~\ref{sec:case_study}. The latter is of particular interest when using our model in practice as it gives the analyst
a refined view on the future reporting times (at daily level or aggregated e.g.~by future reporting weeks or months). An out-of-time evaluation is then a valuable tool to assess the predictive performance of a proposed nowcasting model. Hereby, we restrict the time window, say to events with occurrences $1\leq t \leq \tau^{\star} < \tau$ and reporting delays $d\geq 0$ and $t+d \leq \tau^{\star}$, calibrate the model with the adjusted evaluation date $\tau^{\star}$ and use it to nowcast the events with $1\leq t \leq \tau^{\star}$ and $\tau \geq t+d > \tau^{\star}$. These nowcasts can then be compared to the actual observed reporting of events. We are particularly in favor of performing a moving window out-of-time evaluation, where we repeat the single out-of-time evaluation multiple times with different evaluation dates $\tau^{\star}$. This not only allows to assess the stable predictive performance of a model across multiple evaluation dates, but also allows to verify possible changes in the parameters when calibrated over different subsets of data. When such (substantial) changes are detected, the use of change-points in the modelling of the occurrence and reporting processes may be an interesting direction to explore, see \cite{Tabnak2000} for an example in modelling reporting delays in AIDS surveillance data.

\subsection{Example: revisiting the chain ladder method with an EM algorithm} \label{sec:CL}

The chain ladder introduced in Section~\ref{sec:discrete_time_models} imposes a multiplicative structure on the reporting intensity, with a stationary reporting delay distribution. Classic ways to estimate this model either rely on a log-linear Poisson regression model or use the development factors discussed in the initialization step of an EM algorithm (see Section~\ref{sec:EM_reservingModel}). We now revisit the estimation of the chain ladder method and discuss how its parameters
can equivalently be estimated using our proposed estimation framework.
The complete log-likelihood related to the chain ladder method is similar to \eqref{eq:complete_log_likelihood} with the difference that the reporting delay probabilities $p_d$ do not depend on the occurrence period $t$ and the sum over $d$ runs until $\tau -1$. The latter is linked to the fact that the chain ladder method does not allow for extrapolation beyond the range of data \citep[cfr.][for extensions of the classical chain ladder method which involve the estimation of tail factors]{EnglandVerrall2002}.
The E-step is the same as \eqref{eq:Nk} with, for $t = 1, \ldots, \tau$ and  $d = 0, \ldots, \tau-1$, $p_{td}^{(k-1)} $ replaced by $p_{d}^{(k-1)} $, whereas
the M-step simplifies to
\[  \lambda_t^{(k)}  =  N_{t}^{(k)} = \sum_{d=0}^{\tau - 1}  N_{td}^{(k)}	
\mbox{ and} \quad p_{d}^{(k)}  = \frac{\sum_{t = 1}^{\tau} N_{td}^{(k)} }{ \sum_{t = 1}^{\tau} \sum_{d=0}^{\tau - 1} N_{td}^{(k)} } = \frac{\sum_{t = 1}^{\tau} N_{td}^{(k)} }{ \sum_{t = 1}^{\tau} N_{t}^{(k)} }. \]


In case the chain ladder factors \eqref{eq:chainladder_factors} are used in the initial step (see Section \ref{sec:EM_reservingModel}), such that in fact $N_{td}^{(0)}  = \widehat{\lambda}_t   \,  \widehat{p}_{d}$ for $t = 2, \ldots, \tau$ and $d = \tau - t + 1, \ldots, \tau - 1$ due to the equivalence with the Poisson model, convergence is reached immediately the first time we apply the M-step above. Indeed, we then obtain
\begin{align*}
\lambda_t^{(0)}  &= \sum_{d=0}^{\tau - 1}  N_{td}^{(0)}
= \sum_{d=0}^{\tau - t}  N_{td} + \sum_{d=\tau - t + 1}^{\tau - 1} N_{td}^{(0)}
=  \widehat{\lambda}_t  \sum_{d=0}^{\tau - 1} \widehat{p}_{d}  = \widehat{\lambda}_t  ,
\end{align*}
where we used \eqref{eq:CL_i}. Similarly, using \eqref{eq:CL_j}, we find
\begin{align*}
p_{d}^{(0)}  &= \frac{\sum_{t = 1}^{\tau}  N_{td}^{(0)} }{ \sum_{t = 1}^{\tau} \sum_{d=0}^{\tau - 1} N_{td}^{(0)} }  =  \frac{\sum_{t = 1}^{\tau - d}  N_{td} + \sum_{t = \tau - d + 1}^{\tau}  N_{td}^{(0)}  }{ \sum_{t = 1}^{\tau} \widehat{\lambda}_t  }   =
\frac{\sum_{t = 1}^{\tau} \widehat{\lambda}_t   \,  \widehat{p}_{d}}{ \sum_{t = 1}^{\tau} \widehat{\lambda}_t  }
=    \widehat{p}_{d} .
\end{align*}

\noindent Using an EM algorithm, the iterative steps are easy and intuitive and, upon convergence, the same parameter estimators for $\lambda_t$ and $p_{d}$ are obtained compared to direct maximum likelihood optimization.
When structuring the occurrence and reporting delay parameters (as we propose in Section~\ref{sec:flex_models_discrete_time}), the model can no longer be solved analytically nor formulated as a generalized linear model. An EM algorithm then offers an elegant solution.

\section{Case study: insurance nowcasting}
\label{sec:case_study}
\label{sec:data_insights}

We analyze a data set with the occurrence and reporting dates of claims in a portfolio of general liability insurance policies for private individuals from a European insurance company. The goal is to predict (or: nowcast) the number of claims that already incurred in the past, but are not yet reported to the insurance company.

\subsection{Description of the insurance dataset}

Detailed claim and policy information is available from January 2000 until August 2009. This includes the occurrence date of a claim, the time between occurrence and reporting of the claim to the insurance company and an exposure-to-risk measure. 
The online supplement further details the exposure measure that is available in this dataset. 
To enable out-of-time prediction, we restrict our analysis to claims that have occurred between January 1, 2000 and August 31, 2004.
We set the new evaluation date, say $\tau^{\star}$, at the end of this time window, on August 31, 2004 and want to estimate the total incurred but not reported (IBNR) claim count, as well as the reporting dates of these IBNR claims.
Based on the full data set until August 2009, \num{176671} claims have occurred during this time window.
Due to a reporting delay, only \num{174624} of these have been reported by the evaluation date,
\tcr{as visualized in Figure~\ref{fig:daily_triangle} in the online supplement.}
The remaining \num{2047} claims are IBNR claims, i.e.~claims which have occurred between January 2000 and August 2004 but have only been reported after the evaluation date and before the end of the observation period.

\subsection{Parametric models for the occurrence and reporting processes} \label{section:repdelaymodel}

In line with Section~\ref{sec:flex_models_discrete_time}, we work in discrete time and explicitly model the occurrence and reporting processes.

\paragraph{Occurrence model.} Let $N_{t}$ be the total insurance claim counts that occur on day $t$, for $t=1,\ldots,\tau^{\star}$ where $t=1$ is Jan 1, 2000 and $t=\tau^{\star}$ refers to August 31, 2004. We model the $N_t$ as independent Poisson distributed random variables with intensity $\lambda_t$, see assumption~\ref{A1}, structured as
\begin{eqnarray}\label{eq:occ}
\lambda_t &=& e_t \cdot \exp{\left( \alpha_0 + \alpha_{\texttt{jan1}(t)} + \alpha_{\texttt{dec31}(t)} + \alpha_{\texttt{month}(t)} + \alpha_{\texttt{dow}(t)} + \alpha_{\texttt{dom}(t)}   \right)},
\end{eqnarray}
where $e_t$ is the exposure on day $t$, $\texttt{month}(t)$ indicates the month, $\texttt{dow}(t)$ the day of the week and $\texttt{dom}(t)$ the day of the month to which $t$ belongs. We also include parameters for occurrence days $t$ on January 1 and December 31, because on these days many claims occur.

\paragraph{Reporting model.}

Figures~\ref{fig:reporting_Monday} and \ref{fig:reporting_week} in the online supplement illustrate how reporting probabilities decrease over time and are low on Saturdays and Sundays. Supported by these empirical findings we structure reporting delay as the product of week probabilities and day (or intra-week) probabilities
$
p_{td}(\bs{\theta},\bs{x}_{td}) = p^{\W}_{tw}(\bs{\theta},\bs{x}_{t})  \cdot  p^{\text{intra}}_{td}.
$
Here,
$p^{W}_{tw}$ denotes the probability of reporting an event from occurrence day $t$ in the $w$th week after occurrence. The
intra-week reporting probabilities are denoted with $p^{\text{intra}}_{td}$ and sum to 1 over the days within the reporting week.

%

The reporting week probabilities
\begin{equation*}
p^{\W}_{tw}  =  \frac{ \Gamma(\phi+ w)}{ w! \Gamma(\phi)}    \frac{\phi^{\phi}  \mu_{t}^w }{ (\phi + \mu_{t}) ^{\phi + w} } \quad \mbox {for } w = 0, 1, 2, \ldots  \ ,
\label{eq:pW}
\end{equation*}
are modeled using the probability mass function of a negative binomial distribution with expected value $\mu_{t} = \exp(\bs{x}_t' \bs{\theta})$ and variance $\mu_{t} + \mu_{t}^2/\phi$, where $\phi$ is the dispersion parameter and $\bs{x}_t$ is the covariate vector corresponding to occurrence day $t$. \tcr{Figure~\ref{fig:reporting_week} in the supplement indicates that the negative binomial distribution is a good fit.} 
We structure the $\mu_t$ as follows,
\begin{eqnarray}\label{eq:mupW}
\mu_t &=& \exp{\left( \theta_0 + \theta_{\texttt{jan1}(t)} + \theta_{\texttt{dec31}(t)} + \theta_{\texttt{month}(t)} + \theta_{\texttt{dow}(t)} + \theta_{\texttt{dom}(t)}\right)},
\end{eqnarray}
including parameters for occurrence day $t$ on January 1 and December 31.

A first modelling strategy for the reporting delay day probabilities
symbolically writes these probabilities as
\begin{equation}
p^{\text{intra}}_{td} = \P(\dow(t), \wday(t, t+d)) \, ,
\label{eq:P}
\end{equation}
where $\wday(t, t+d)$ orders the working days within the reporting week with separate levels for Saturday and Sunday. For example, when the claim occurred on a Thursday,
Friday is $\textsf{wday2}=\wday(t,t+1)$ and Monday is $\textsf{wday3}=\wday(t,t+2)$.
The $7 \times 7$-matrix $\P$ then contains the day probabilities related to the first week. Each element in $\P$ is between 0 and 1 and
all row sums equal 1.

Alternatively, we project the seven intra-week reporting probabilities $p^{\text{intra}}_{td}$ to six probabilities, 
inspired by the reverse time strategy of \cite{KalbfleischLawless1991}. This allows to leave the sum-to-one restriction on the intra-week reporting probabilities. Moreover, we can now include covariate information related to both the day of occurrence $t$ and the actual reporting date $t+d$. We define
\begin{align}
	q_{t, 7 w + j} &= P(\text{delay} = 7 w + j \, | \, 7 w \leq \text{delay} \leq 7 w + j) \nonumber \\
	&=\frac{p_{t,7 w + j}^{\text{intra}}}{\sum_{k=0}^{j} p_{t,7 w + k}^{\text{intra}}} \quad \mbox {for } j = 1, \ldots, 6, \label{eq:qprob}
\end{align}
with $w =  \left\lfloor \frac{d}{7} \right\rfloor$ (the reporting week after occurrence) and $j = d - 7w$  the intra-week day of reporting. Expression \eqref{eq:qprob} takes the form of a discrete time hazard rate, but due to the reverse time strategy we now condition on failure before day $7w + j$ instead of survival.
We structure
\begin{equation}\label{eq:q_probs}
\texttt{logit}(q_{t, d}) = \boldsymbol{x}^{'}_{td}\boldsymbol{\gamma} = \gamma_0 + \gamma_{\texttt{workdays}(t, t+d)} + \gamma_{\texttt{dow}(t+d)} + \gamma_{\texttt{holiday}(t+d)},
\end{equation}
where $\texttt{workdays}(t, t+d)$ counts the number of elapsed working days in the current reporting week, i.e.~excluding the weekend and holidays,
and $\texttt{holiday}(t + d)$ indicates whether $t+d$ is a holiday. Reporting on holidays is exceptional in our data set.
The inclusion of dummy variables for holidays allows to capture this empirical fact in the intra-day reporting probabilities, while distinguishing between national holidays on which all companies are closed and two unofficial holidays (New Year's Eve and Good Friday). 

\subsection{Competing strategies: directly modelling the reporting intensity and the chain ladder method}\label{sec:ladder}

We also study some competing modelling strategies proposed in the literature.
We consider two versions of the models discussed in Section~\ref{sec:discrete_regr} where the $N_{td}$ are independent Poisson distributed random variables with mean $\lambda_{td}$ structured as
\begin{eqnarray}\label{eq:rep_int_1}
\lambda_{td} &= e_t \cdot \exp\big(\alpha_{\texttt{jan1}(t)} + \alpha_{\texttt{dec31}(t)} + \alpha_{\texttt{month}(t)} + \alpha_{\texttt{dow}(t)} + \alpha_{\texttt{dom}(t)} + \nonumber \\
  &\phantom{{}={}}\beta_{\texttt{holiday}(t+d)} + \beta_{\texttt{holiday}(t+d+1)} + \beta_{\texttt{dow}(t+d)} + \beta_{\texttt{delay}(d)} \big),
\end{eqnarray}
or
\begin{eqnarray}\label{eq:rep_int_2}
\lambda_{td} &= e_t \cdot  \exp\big(\alpha_t + \beta_{\texttt{holiday}(t+d)} + \beta_{\texttt{holiday}(t+d+1)} + \nonumber \\
  &\phantom{{}={}}\beta_{\texttt{dow}(t+d)} + \beta_{\texttt{delay}(d)} \big).
\end{eqnarray}

While \eqref{eq:rep_int_1} uses covariates capturing relevant information from $t$, \eqref{eq:rep_int_2} includes a parameter $\alpha_t$ for each $t$. Both model specifications structure the information on the reporting delay with a parameter $\beta_{\texttt{delay}(d)}$ for each observed delay \citep[see e.g.][]{Bastos} as well as $\beta_{\texttt{dow}(t+d)}$ referring to the day of the week of the effective reporting date $t+d$. The  $\beta_{\texttt{holiday}(t+d)}$ and $\beta_{\texttt{holiday}(t+d+1)}$ correspond to reporting on a  holiday, or the day after. No explicit reporting delay distribution is proposed, but instead a large number of parameters is used. 

We also calibrate the chain ladder method using a yearly grid. That is,
\begin{eqnarray}\label{eq:chainladder_yearly}
\lambda_{td} &=& \lambda_{\texttt{year(t)}} \cdot p_{\texttt{year}(t+d) - \texttt{year}(t)},
\end{eqnarray}
where $\lambda_{\texttt{year(t)}}$ is the effect of the year of occurrence to which day $t$ belongs. Reporting delay is modelled independently from the occurrence, with a parameter per year of reporting, thus $p_{\texttt{year}(t+d) - \texttt{year}(t)}$ describes the effect of the reporting year of the claims that occurred on day $t$ and were reported with $d$ days of delay.

\subsection{Parameter estimates}

For the models proposed in Section~\ref{section:repdelaymodel}, we use the EM algorithm of Section~\ref{sec:EM} to estimate the parameters in the occurrence and reporting processes. For the  models in Section~\ref{sec:ladder}, maximum likelihood estimates are obtained with
the glm4 routine in the R package MatrixModels \citep{MatrixModels}.

To grasp the insights obtained via an explicit specification of the occurrence and reporting processes, we focus on the maximum likelihood estimates of the parameters in the occurrence model \eqref{eq:occ} combined with the reporting model with reporting week probabilities in \eqref{eq:mupW} and intra-week reporting probabilities in \eqref{eq:P}. We evaluated the impact of the censoring term in the reporting delay likelihood used in the M-step, see \eqref{eq:Q_theta}. We obtained similar parameter estimates when omitting the reporting of events with a delay of more than $\tau^{\star}-1$ days compared to the estimates obtained under the simplified likelihood that drops the censoring term in \eqref{eq:Q_theta} (comparison not shown). Therefore, the results  below are obtained with the simplified reporting likelihood. \tcr{The parameter estimates of some alternative specifications are deferred to Section~\ref{app:add_parm} in the online supplement.}

\begin{figure}[t!]%
\centering
\includegraphics[width=\columnwidth]{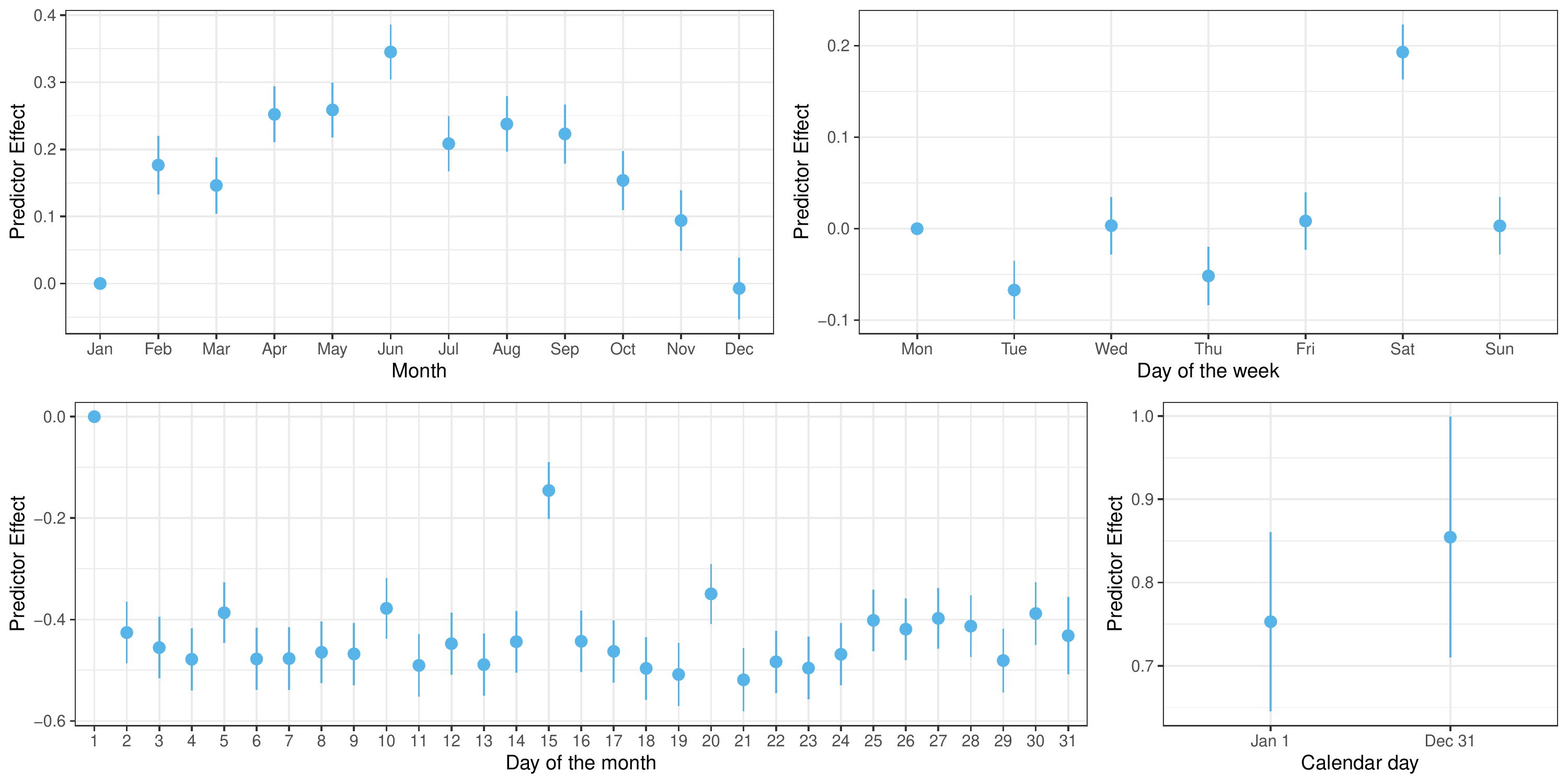}%
\caption{Maximum likelihood estimates and 95\% simultaneous confidence intervals for $\bs{\alpha}$
in the Poisson claim occurrence model.  }%
\label{fig:occurrenceModel}%
\end{figure}

The effects related to the categorical predictors in the Poisson occurrence model specified in \eqref{eq:occ} are visualized in Figure \ref{fig:occurrenceModel}. \tcr{Figure \ref{fig:reportingModel} in the online supplement shows the parameter estimates for the negative binomial regression model of the reporting delay in weeks, while Table~\ref{tab:P} in the online supplement collects the estimated intra-week probabilities from \eqref{eq:P}.} The maximum likelihood estimates are shown along with Bonferroni adjusted simultaneous 95\% confidence intervals based on the inverse of the expected information matrix. The intercept in the Poisson model is estimated as -5.965 with 95\% confidence interval $[-6.020, -5.910]$.

Figure~\ref{fig:occurrenceModel} reveals a seasonal pattern in which the number of claims rises in the middle of the year and falls around the year end. Claims most likely occur in June and least likely in December with an estimated difference in expected value of 42\%. The calibrated day of the week effect shows an increase in the expected number of claims on Saturdays and a slight decrease on Tuesdays and Thursdays. The categorical effect of the day of the month shows a remarkable pattern which is similar in both the claim occurrence and the reporting delay model. On the 1st and 15th, the number of claims as well as the reporting delays have significantly higher expected values. To a lesser extent, this is also present for the 5th, 10th, 20th, 25th, and 30th or 31st day of each month. This pattern can most likely be explained by rounding errors of the occurrence date when insureds have to report a claim which took place several weeks or months ago. Since this misreporting of dates is more likely to occur for claims which are only reported after a longer time period, we simultaneously see an increase in the expected reporting delay for claims occurring on these rounded month days.
There is a high number of occurrences on December 31 and January 1 with on average 112\%, respectively 135\%, more claims. However, the reporting delay of these claims is not significantly different from the other ones.


\subsection{Prediction of unreported claim counts}\label{sec:pred_detail} 

The main goal of our proposed models is to estimate the number of unreported claims. Using an out-of-time evaluation with $\tau^{\star} = $ August 31, 2004, we know there are \num{2047} IBNR claims in the full data set, which runs until August 2009, of which we know the corresponding occurrence date and reporting delay.

\begin{figure}[htb!]%
\centering
 \begin{subfigure}[t]{\textwidth}
          \centering
					\caption{}%
					\vspace{-0.2cm}
					\includegraphics[width=\columnwidth]{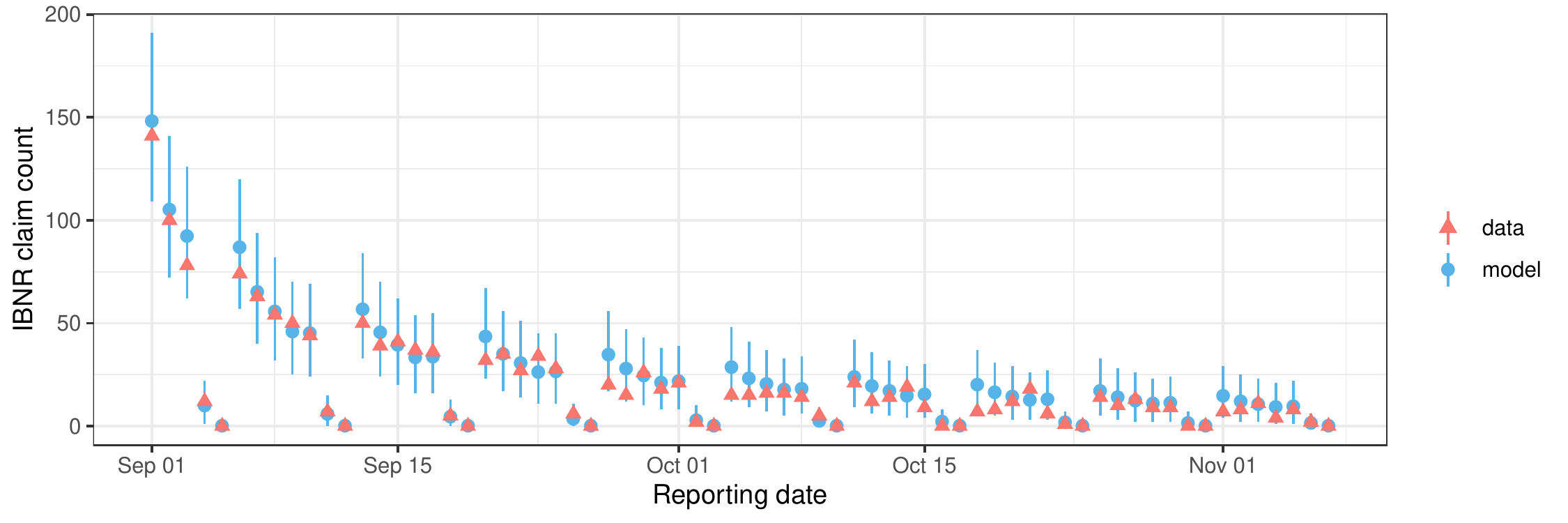}%
					\label{fig:IBNR_reporting_day}%
					\vspace{-0.4cm}
     \end{subfigure}
		   \begin{subfigure}[t]{0.5\textwidth}
          \centering
					\caption{}%
					\vspace{-0.2cm}
					\includegraphics[width=\columnwidth]{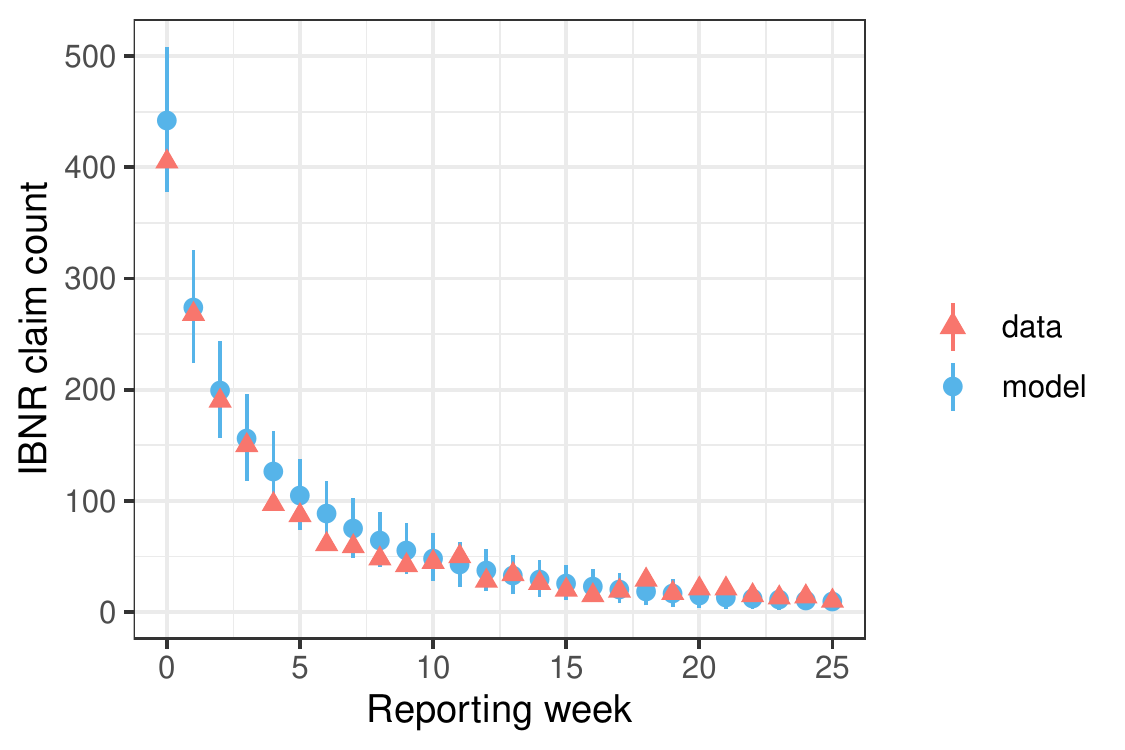}%
					\label{fig:IBNR_reporting_week}%
     \end{subfigure}%
		 \begin{subfigure}[t]{0.5\textwidth}
          \centering
					\caption{}%
					\vspace{-0.2cm}
					\includegraphics[width=\columnwidth]{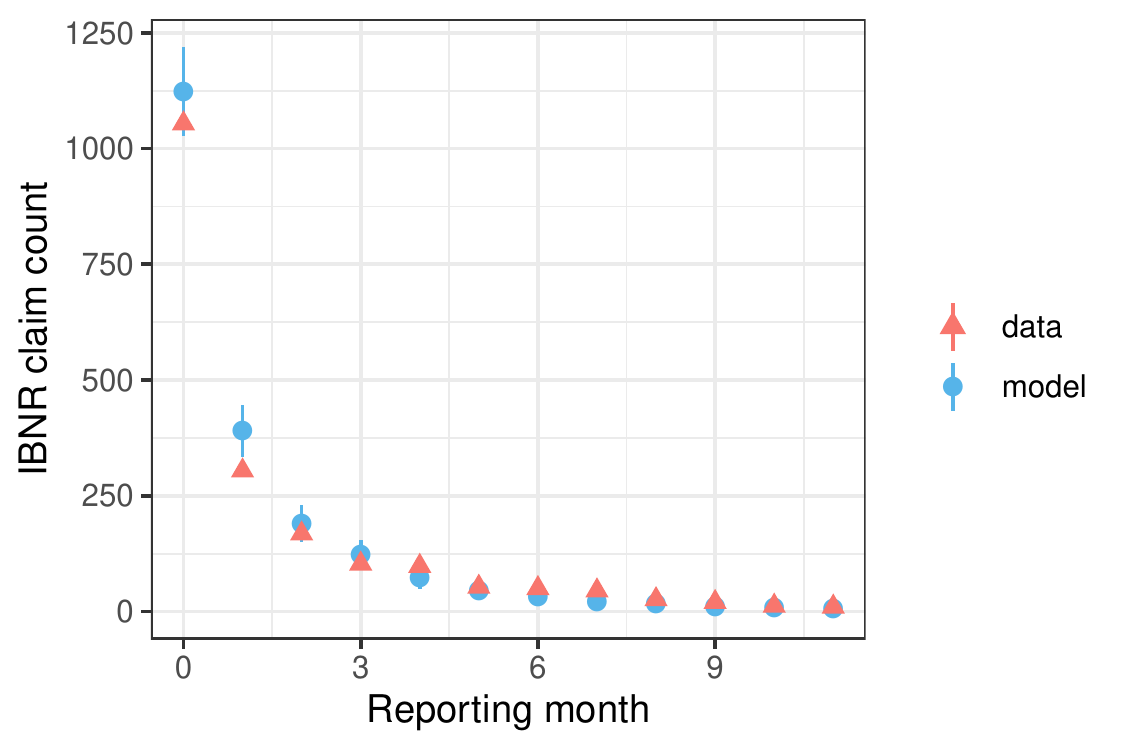}%
					\label{fig:IBNR_reporting_month}%
     \end{subfigure}
\caption{Predictions of the IBNR claim counts and simultaneous 95\% prediction intervals by reporting date.
(a) Daily, for reporting dates in between September 1 and November 7, 2004; (b) weekly (7 days) for the next 26 weeks; (c) monthly (30 days) for the next 12 months.}
  \label{fig:IBNR_reporting}  %
\end{figure}

We first demonstrate the insights that can be derived from a nowcasting model using the occurrence specification in \eqref{eq:occ} and the reporting model in \eqref{eq:mupW} combined with \eqref{eq:P}. Calibrated on the data from January 1, 2000 to August 31, 2004 this model estimates the total number of IBNR claims as $\widehat{N}^{\IBNR} = 2055.8$, which is close to the actual count. Moreover, the distributional assumptions of our model can be used to provide a 95\% prediction interval given by $[1697, 2145]$, see Section \ref{sec:prediction}.
Furthermore, since the model is defined on a daily level, the total IBNR prediction can be divided into daily forecasts by occurrence date and by reporting date.
To illustrate this strong point of our model, we predict the IBNR claim counts by reporting date in Figure \ref{fig:IBNR_reporting}. \tcr{Figure~\ref{fig:IBNR_occurrence} in the online supplement shows a similar split of the IBNR claims by occurrence dates.}

In Figure \ref{fig:IBNR_reporting} we disperse the total predicted IBNR claim count according to the date on which the claims will be reported to the insurer.
It means we now focus on estimating $\sum_{t = 1}^{\tau^{\star}}  N_{t, \rho-t}$ for $\rho = \tau^{\star} + 1, \tau^{\star}+2, \ldots$, i.e.~the number of unreported claims reported on day 1, 2, \ldots of the out-of-time period.
This forms an appealing way to use our model in practice as it gives the insurer a refined view on the reporting times.
The predictions on a daily level in Figure \ref{fig:IBNR_reporting_day} are accompanied by 95\% simultaneous prediction intervals and range from September 1, 2004, until November 7, 2004, i.e.~the first two months following our training period.
When compared to the out-of-time actual values, the forecasts clearly capture the downward trend in the reporting of IBNR claims and the nearly absence of reporting in weekends.
This is primarily the case due to the day probabilities in our model which reflect the day-specific aspects of the reporting delay.
In Figure \ref{fig:IBNR_reporting_week} (resp.~\ref{fig:IBNR_reporting_month}) the reporting dates are grouped by weeks (resp.~months) after the evaluation date and the IBNR claim counts are predicted for the next 26 weeks (resp.~12 months).
We notice how, also over longer time spans, the predictions by reporting week or month follow the pattern observed in the actual unreported counts.

\subsection{Comparing nowcasting models with a moving window evaluation}\label{sec:model_comparison}

We compare the performance of different model specifications via a
moving window out-of-time evaluation. We adjust the evaluation date $\tau^{\star}$, which was previously chosen to be August 31, 2004, to any date in between August 31, 2003, and August 31, 2004. For each such $\tau^{\star}$, we refit the model based on the observed data by that date, $\bs{N}^{\r} = \{  N_{td} \mid 1 \leq t \leq \tau^{\star}, d \geq 0, t + d \leq \tau^{\star} \}$, and produce an estimate of the total unreported count $N^{\IBNR} = \sum_{t = 1}^{\tau^{\star}} N_{t}^{\IBNR}$ corresponding to claims that occurred before or at $\tau^{\star}$.

\begin{figure}[htb!]%
\centering

\begin{subfigure}[t]{0.5\textwidth}
          \centering
					\caption{}%
					\vspace{-0.2cm}
					\includegraphics[width=\columnwidth]{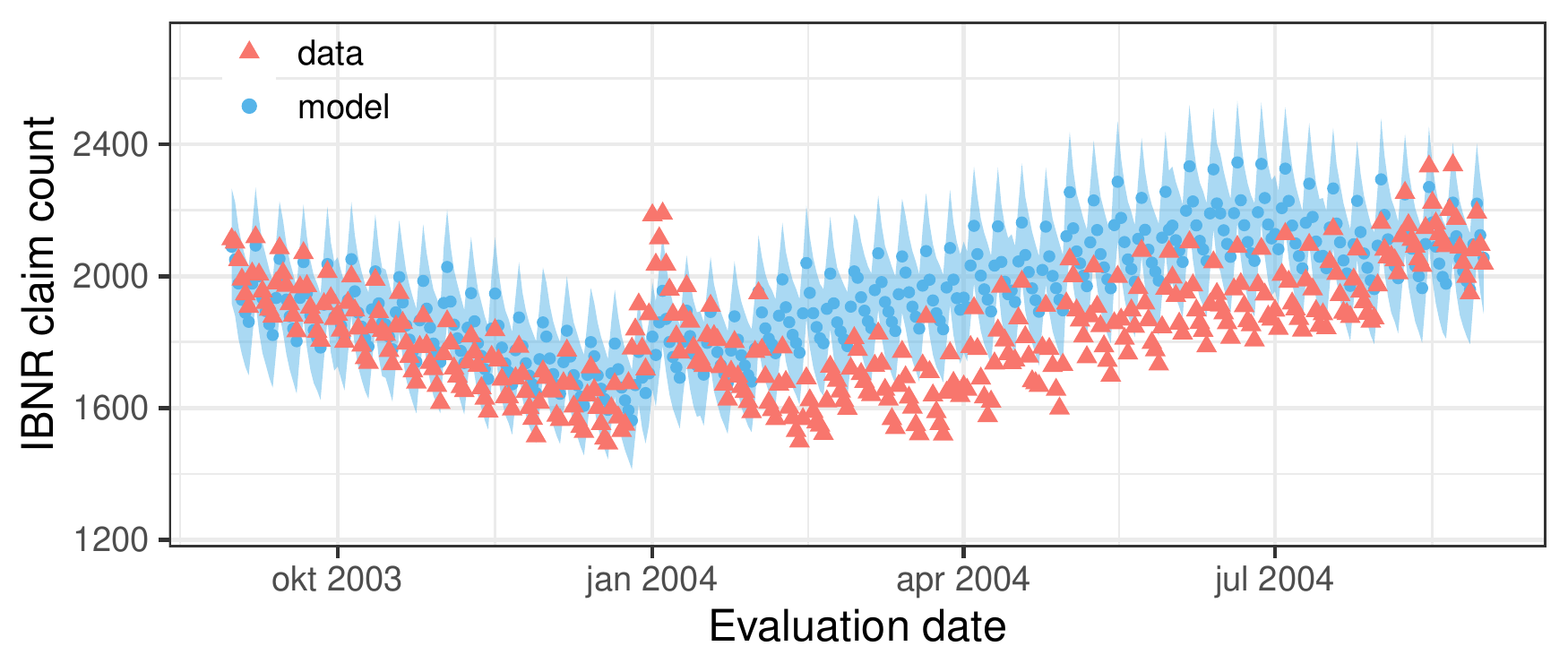}%
					\label{fig:IBNR_daily_total_em_pd_full}%
     \end{subfigure}%
		 \begin{subfigure}[t]{0.5\textwidth}
          \centering
					\caption{}%
					\vspace{-0.2cm}
					\includegraphics[width=\columnwidth]{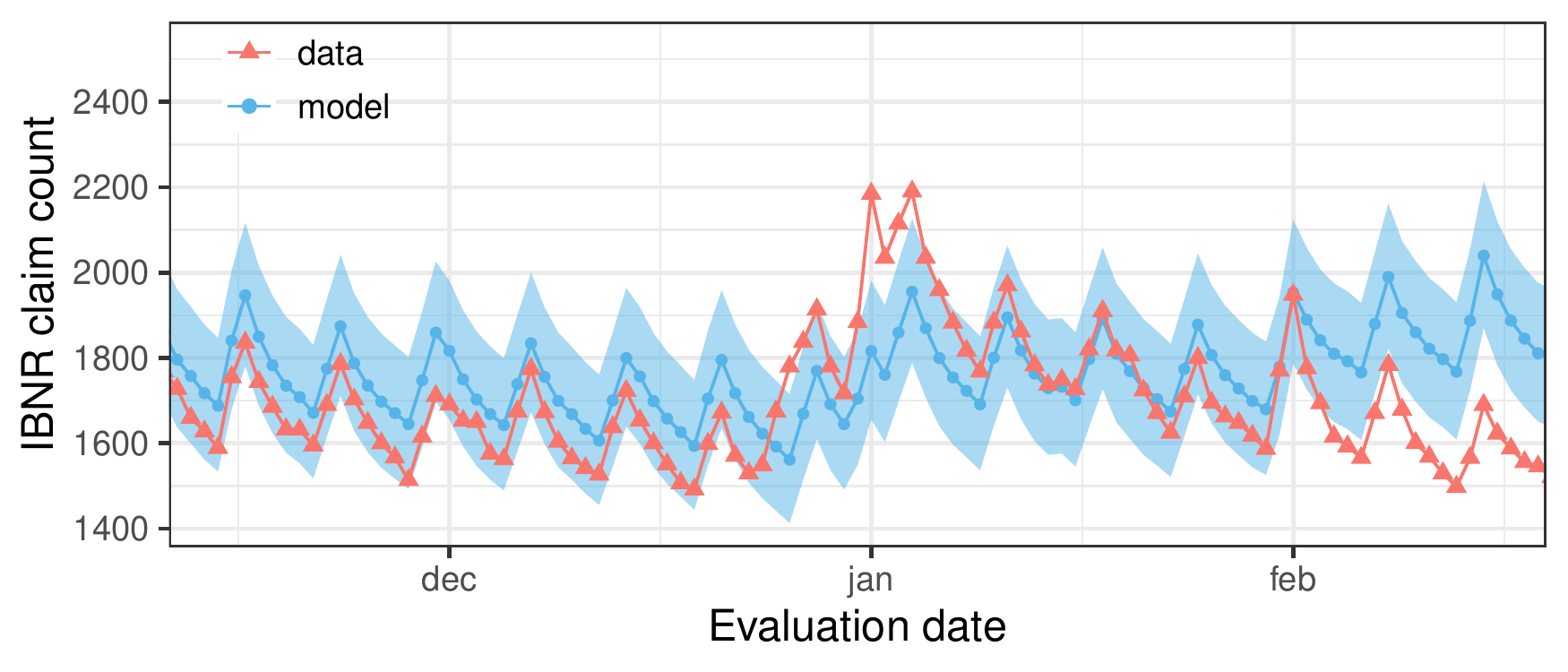}%
					\label{fig:IBNR_daily_total_em_pd_zoom}%
     \end{subfigure}

\caption{Predictions and 95\% simultanteous prediction intervals of the total IBNR claim counts for varying evaluation dates $\tau^{\star}$: (a) in between August 31, 2003, and August 31, 2004 and (b) zoom in on the estimates from November 15 to February 15. Red triangles: data, blue circles: model.
}

\label{fig:IBNR_daily_total_em_pd}%
\end{figure}

Figures~\ref{fig:IBNR_daily_total_em_pd_full} and (zoom in) \ref{fig:IBNR_daily_total_em_pd_zoom} show the predictions obtained with the occurrence specification in \eqref{eq:occ} and the reporting model in \eqref{eq:mupW} and \eqref{eq:P}, as well as the actual total IBNR claim counts based on the full data set.
Overall, the estimates follow the seasonal pattern quite well. The deviations from the observed trend in the first months of 2014 might be explained by variations from year to year in the effect of the month on the occurrence and reporting of claims. In
our model, we assume the seasonal monthly pattern to be the same over the different years. As a result, the parameter estimates are averaged values. The model could potentially be refined (by including extra covariates) if expert-knowledge is available on how changes in e.g.~product design and conditions, the claims handling process, the business environment or legislation, may impact the claim arrival process or the reporting delay.
We notice a seven day pattern in the number of unreported claims in Figure~\ref{fig:IBNR_daily_total_em_pd_zoom}, which results from the delayed reporting during the weekend. This aspect is incorporated in our model through the intra-week reporting probabilities in \eqref{eq:P}. Moreover, we observe an increase in the number of IBNR claim counts around the end of the year. This is due to the fact that the insurance company is closed around the holidays, preventing any claims from being reported at that time. This effect is only partially captured by the dummy variables for the occurrence dates \texttt{jan1} and \texttt{dec31} in \eqref{eq:occ} and \eqref{eq:mupW}.

\begin{figure}[htb!]%
\centering

	\begin{subfigure}[t]{0.5\textwidth}
          \centering
					\caption{}%
					\vspace{-0.2cm}
					\includegraphics[width=\columnwidth]{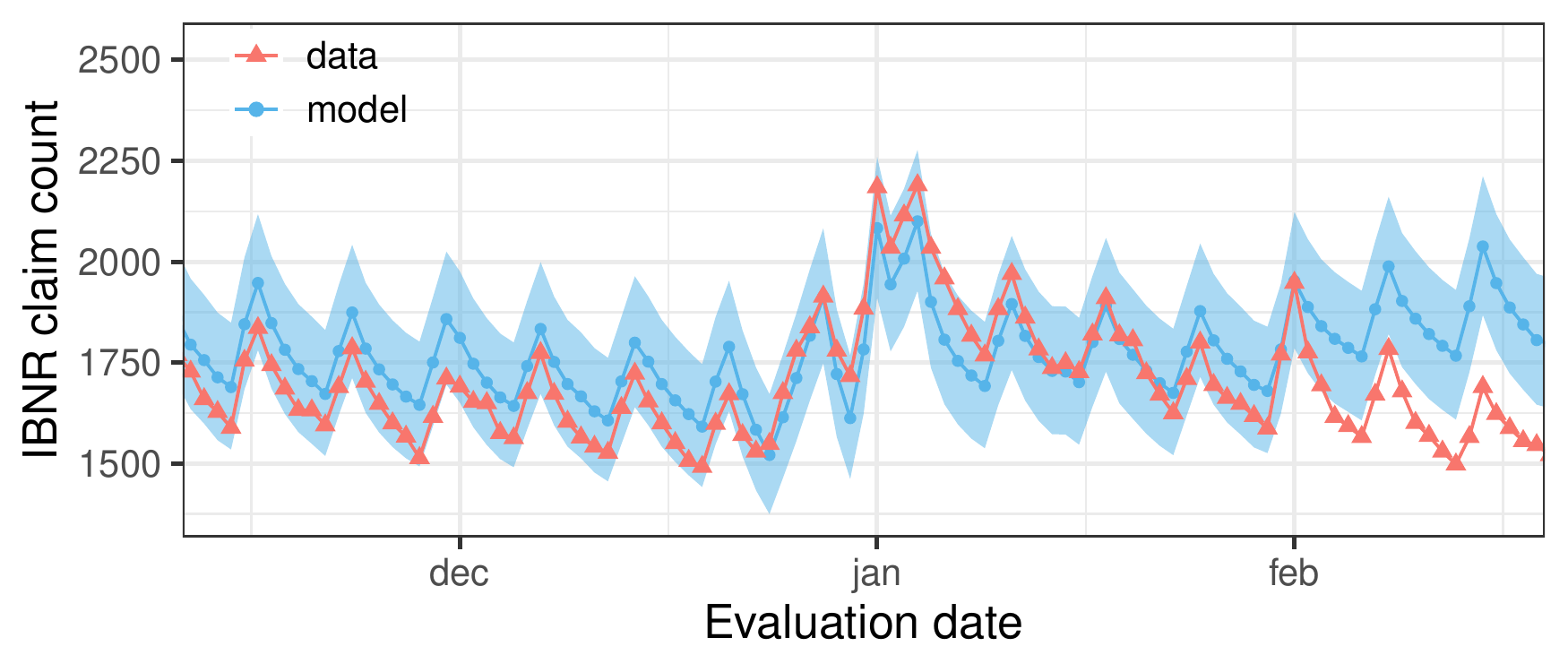}%
					\label{fig:ibnr_evolution_holiday_zoom}%
     \end{subfigure}%
		 \begin{subfigure}[t]{0.5\textwidth}
          \centering
					\caption{}%
					\vspace{-0.2cm}
					\includegraphics[width=\columnwidth]{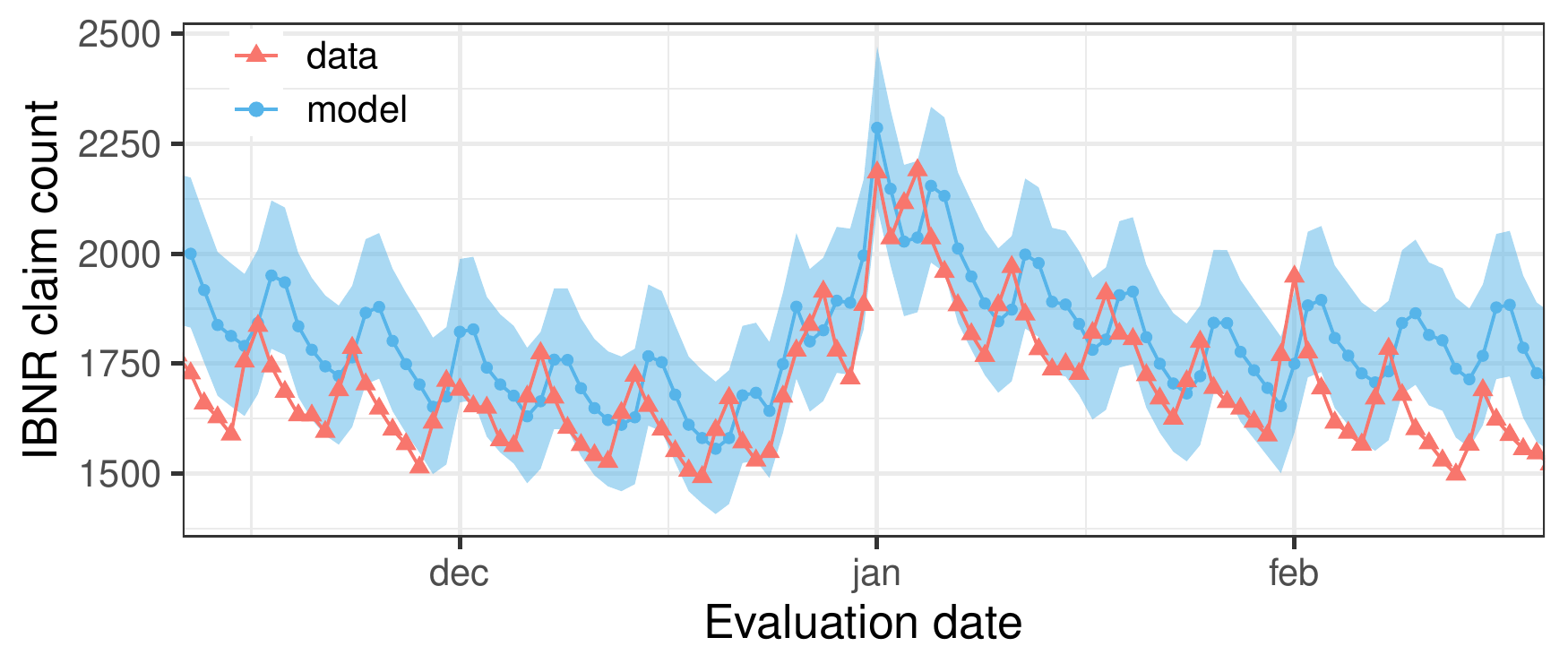}%
					\label{fig:ibnr_evolution_cl_zoom}%
     \end{subfigure}

     \begin{subfigure}[t]{0.5\textwidth}
          \centering
					\caption{}%
					\vspace{-0.2cm}
					\includegraphics[width=\columnwidth]{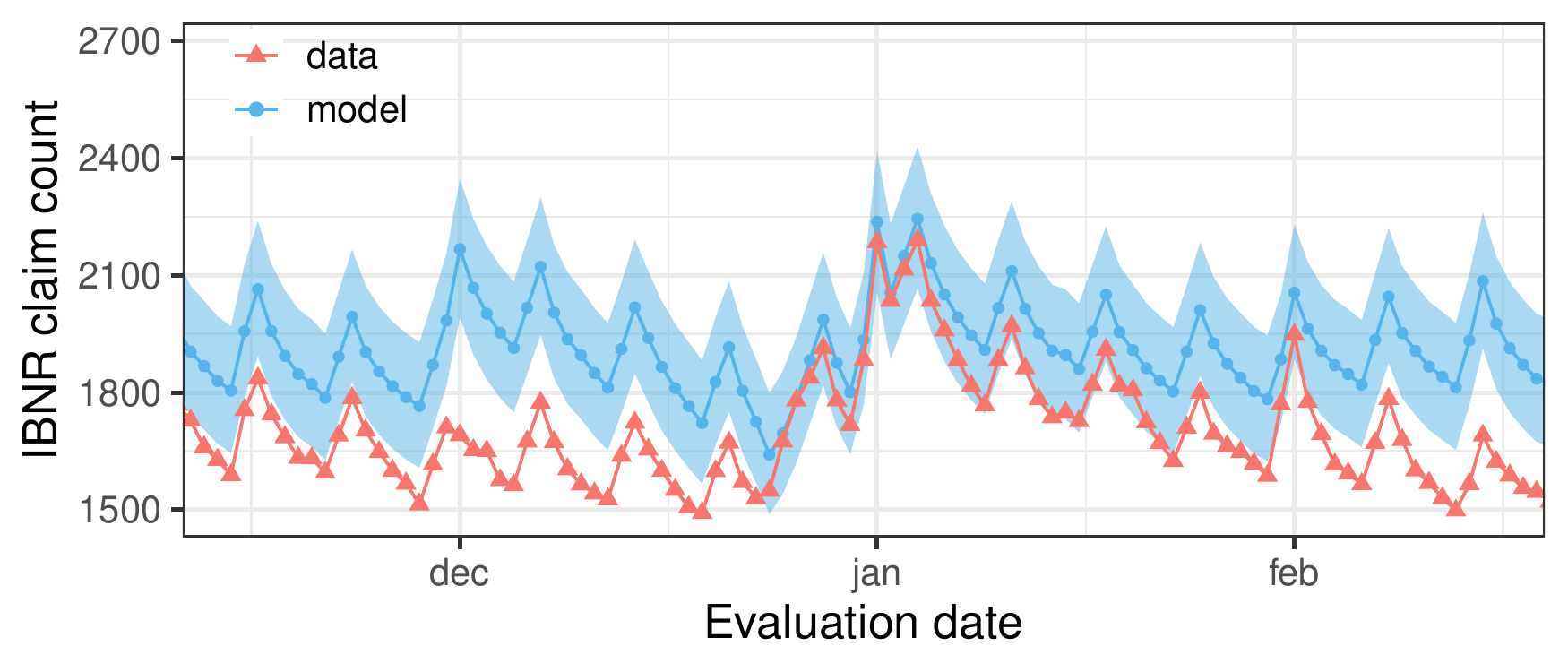}%
					\label{fig:ibnr_evolution_nowcasting_zoom}%
     \end{subfigure}%
		 \begin{subfigure}[t]{0.5\textwidth}
          \centering
					\caption{}%
					\vspace{-0.2cm}
					\includegraphics[width=\columnwidth]{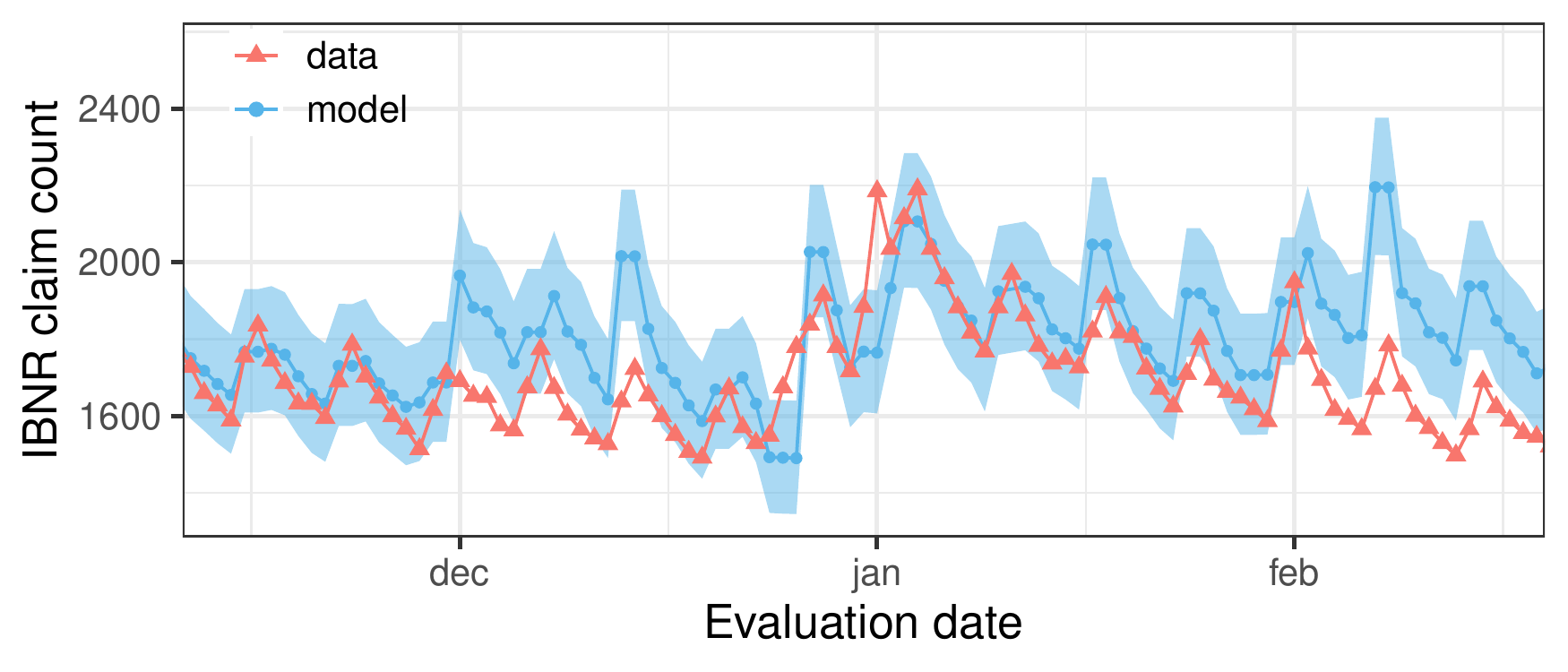}%
					\label{fig:ibnr_evolution_nowcasting_occ_zoom}%
     \end{subfigure}

\caption{Predictions and 95\% simultaneous prediction intervals of the total IBNR claim counts for varying evaluation dates from November 15 to February 15, 2004. (a) uses \eqref{eq:occ}, \eqref{eq:mupW} and \eqref{eq:q_probs}, (b) yearly chain ladder method specified in \eqref{eq:chainladder_yearly},
(c) direct specification of reporting intensity in \eqref{eq:rep_int_1}, (d) direct specification of reporting intensity in \eqref{eq:rep_int_2}.
Red triangles: data, blue circles: model.}%
\label{fig:IBNR_compared-zoom}%
\end{figure}

We compare these predictions to two other model specifications that benefit from an EM algorithm for calibration.
Figure~\ref{fig:IBNR_compared-zoom} zooms in on evaluation dates between November 15 and February 15, 2004. \tcr{The corresponding plots with evaluation dates between August 31, 2003 and August 31, 2004 are available in the online supplement, Figure~\ref{fig:IBNR_compared-full}.} Figure~\ref{fig:ibnr_evolution_holiday_zoom} shows predictions obtained with the occurrence specification in \eqref{eq:occ} and the reporting model in \eqref{eq:mupW} and \eqref{eq:q_probs}. In the specification of the intra-week probabilities this model designates dummy indicators for reporting on national and unofficial holidays in \eqref{eq:q_probs}, which leads to clear improvements around the end-of-year holidays as compared to Figure~\ref{fig:IBNR_daily_total_em_pd}.
The yearly chain ladder method is used in Figure~\ref{fig:ibnr_evolution_cl_zoom}.
While this method overall performs well, detailed insights on the number of reportings in e.g.~the next days, weeks or months can not be deduced from this model calibrated on data aggregated into yearly grids. Moreover, each year ends on a different day of the week. Since the chain ladder averages the reporting probability over past years, the estimated 7-day week pattern in the predictions is slightly misaligned with the actual pattern in the data. \cite{crevecoeur2019} show that when more years of data are available the 7-day pattern completely disappears from the yearly chain ladder predictions, which then results in a systematic underestimation of the IBNR claim count on Saturdays and Sundays. Experiments to capture the week pattern with a more granular chain ladder method resulted in a sharp loss in the overall performance (not shown).

The nowcasting models illustrated in Figure~\ref{fig:ibnr_evolution_nowcasting_zoom} and \ref{fig:ibnr_evolution_nowcasting_occ_zoom}
directly specify a structure for the reporting intensity. In \eqref{eq:rep_int_1} (shown in panel \ref{fig:ibnr_evolution_nowcasting_zoom}) the information on the occurrence date $t$ is structured via a set of covariates, whereas panel \ref{fig:ibnr_evolution_nowcasting_occ_zoom} calibrates a parameter for each occurrence date $t$ (as in \eqref{eq:rep_int_2}).
With this type of models it is more difficult to capture insights from the calibrated parameters, as there is no explicit distinction between the dynamics in the occurrence and the reporting process. On the one hand, with a structured representation of the influence of the occurrence dates $t$ does not sufficiently capture the actual pattern in the data. The model specification with a parameter for each occurrence date $t$ on the other hand leads to volatile nowcasts, e.g., one prediction was removed ($\num{96995}$ predicted IBNR claims on Sunday, January 11 2004). Since reporting on a Sunday is extremely rare in our data set the occurrence and immediate reporting of a claim on this date resulted in an unreasonably high $\lambda_t$ estimate.

\section{Conclusions and outlook} \label{sec:conclusions}

We review and structure the literature on nowcasting the occurrence of events subject to a reporting or observation delay. Our literature overview bridges multiple disciplines and incorporates papers from the actuarial, statistical and epidemiological literature. We propose a general modelling and estimation framework capable of dealing with any parametric structure for both the occurrence as
well as the reporting process. The framework uses regression models for count data and treats the right truncation of the reporting delays as a type of missing data. Applying an EM algorithm strongly simplifies maximum likelihood estimation as it allows for the use of standard statistical software to fit the regression models. As an example, we demonstrate how the parameter estimators of a classical method for nowcasting with aggregated data, also known as the chain ladder method, can be obtained using an EM approach.
We investigate the performance of our proposed framework on an insurance case study where the focus is on predicting the number of incurred but not (yet) reported claims in a European portfolio of general liability insurance policies for private individuals. We benchmark the predictions obtained with our proposed strategy against the results from other nowcasting models that directly structure the reporting intensity and can be calibrated with standard statistical routines. The presented model provides a better understanding of the claim occurrence as well as reporting delay process and leads to a refined view on the future reporting times (at daily level or aggregated
e.g. by future reporting weeks or months).

We indicate some possible directions for future research. First of all, we would like to stress that the provided estimation framework involving an EM algorithm is applicable in a wide context (e.g.~epidemiology and reliability engineering) whenever cases are only reported after a delay. This provides a more desirable alternative over the ad hoc methods or two-step approaches used earlier in the literature. The estimation procedure described in Section \ref{sec:EM} is readily applicable to other contexts after specifying a suitable parametric model for the reporting delay probabilities.

It would then be interesting to explore different distributional assumptions for the daily total event counts and the reporting delay structure. The reporting delay can be easily altered within the given framework to, for instance, a zero-inflated or hurdle
distribution or a more heavy-tailed distribution. Relaxing the Poisson assumption for the daily total event counts is also feasible but might complicate the E-step in which we now relied on the thinning property of Poisson distributions. The EM framework is however compatible with latent underlying processes affecting the occurrence of events such as hidden Markov models or shot noise process \citep[see e.g.][]{badescu2016estimation,avanzi2016micro}. Another promising approach would be to investigate how time series models for counts \citep[see][for an overview]{Jung2011} could be introduced in this setting.

\section*{Acknowledgements}

The authors are grateful to the editors, associate editor and the referee for the valuable comments and suggestions.

\bibliographystyle{Chicago}
\bibliography{reserving}

\begin{thebibliography}{}

\bibitem[\protect\citeauthoryear{Akaike}{Akaike}{1973}]{Akaike73}
Akaike, H. (1973).
\newblock Information theory and an extension of the maximum likelihood
  principle.
\newblock In B.~Petrov and F.~Cs\'{a}ki (Eds.), {\em Second {I}nternational
  {S}ymposium on {I}nformation {T}heory}, pp.\  267--281. Akad{\'e}miai
  Kiad\'{o}, Budapest.

\bibitem[\protect\citeauthoryear{Antonio and Plat}{Antonio and
  Plat}{2014}]{AntonioPlat2014}
Antonio, K. and R.~Plat (2014).
\newblock Micro-level stochastic loss reserving for general insurance.
\newblock {\em Scandinavian Actuarial Journal\/}~{\em 7}, 649--669.

\bibitem[\protect\citeauthoryear{Avanzi, Wong, and Yang}{Avanzi
  et~al.}{2016}]{avanzi2016micro}
Avanzi, B., B.~Wong, and X.~Yang (2016).
\newblock A micro-level claim count model with overdispersion and reporting
  delays.
\newblock {\em Insurance: Mathematics and Economics\/}~{\em 71}, 1--14.

\bibitem[\protect\citeauthoryear{Bacchetti, Segal, and Jewell}{Bacchetti
  et~al.}{1993}]{Bacchettietal1993}
Bacchetti, P., M.~R. Segal, and N.~P. Jewell (1993).
\newblock Backcalculation of {HIV} infection rates.
\newblock {\em Statistical Science\/}~{\em 8\/}(2), 82--101.

\bibitem[\protect\citeauthoryear{Badescu, Lin, and Tang}{Badescu
  et~al.}{2016}]{badescu2016marked}
Badescu, A.~L., X.~S. Lin, and D.~Tang (2016).
\newblock A marked {C}ox model for the number of {IBNR} claims: Theory.
\newblock {\em Insurance: Mathematics and Economics\/}~{\em 69}, 29--37.

\bibitem[\protect\citeauthoryear{Badescu, Lin, and Tang}{Badescu
  et~al.}{2019}]{badescu2016estimation}
Badescu, A.~L., X.~S. Lin, and D.~Tang (2019).
\newblock A marked {C}ox model for the number of {IBNR} claims: Estimation and
  application.
\newblock {\em {ASTIN} {B}ulletin\/}~{\em 49}, 709--739.

\bibitem[\protect\citeauthoryear{Barreto-Souza and Simas}{Barreto-Souza and
  Simas}{2017}]{BarretoSouzaSimas2017}
Barreto-Souza, W. and A.~B. Simas (2017).
\newblock Improving estimation for beta regression models via {EM}-algorithm
  and related diagnostic tools.
\newblock {\em Journal of Statistical Computation and Simulation\/}~{\em
  87\/}(14), 2847--2867.

\bibitem[\protect\citeauthoryear{Bastos, Economou, Gomes, Villela, Coelho,
  Cruz, Stoner, Bailey, and Code\c{o}}{Bastos et~al.}{2019}]{Bastos}
Bastos, L.~S., T.~Economou, M.~F. Gomes, D.~A. Villela, F.~C. Coelho, O.~G.
  Cruz, O.~Stoner, T.~Bailey, and C.~T. Code\c{o} (2019).
\newblock A modelling approach for correcting reporting delays in disease
  surveillance data.
\newblock {\em Statistics in Medicine\/}~{\em 38\/}(22), 4363--4377.

\bibitem[\protect\citeauthoryear{Bates and Maechler}{Bates and
  Maechler}{2015}]{MatrixModels}
Bates, D. and M.~Maechler (2015).
\newblock {\em MatrixModels: Modelling with Sparse And Dense Matrices}.
\newblock R package version 0.4-1.

\bibitem[\protect\citeauthoryear{Becker and Cui}{Becker and
  Cui}{1997}]{BeckerCui1997}
Becker, N.~G. and J.-S. Cui (1997).
\newblock Estimating a delay distribution from incomplete data, with
  application to reporting lags for {AIDS} cases.
\newblock {\em Statistics in Medicine\/}~{\em 16\/}(20), 2339--2347.

\bibitem[\protect\citeauthoryear{Bellocco and Marschner}{Bellocco and
  Marschner}{2000}]{Bellocco2000}
Bellocco, R. and I.~C. Marschner (2000).
\newblock Joint analysis of {HIV} and {AIDS} surveillance data in
  back-calculation.
\newblock {\em Statistics in Medicine\/}~{\em 19\/}(3), 297--311.

\bibitem[\protect\citeauthoryear{Brookmeyer and Gail}{Brookmeyer and
  Gail}{1988}]{Brookmeyer1988}
Brookmeyer, R. and M.~H. Gail (1988).
\newblock A method for obtaining short-term projections and lower bounds on the
  size of the {AIDS} epidemic.
\newblock {\em Journal of the American Statistical Association\/}~{\em
  83\/}(402), 301--308.

\bibitem[\protect\citeauthoryear{Cavanaugh and Shumway}{Cavanaugh and
  Shumway}{1998}]{CavanaughShumway1998}
Cavanaugh, J.~E. and R.~H. Shumway (1998).
\newblock An {A}kaike information criterion for model selection in the presence
  of incomplete data.
\newblock {\em Journal of Statistical Planning and Inference\/}~{\em 67},
  45--65.

\bibitem[\protect\citeauthoryear{Claeskens and Consentino}{Claeskens and
  Consentino}{2008}]{ClaeskensConsentino2008}
Claeskens, G. and F.~Consentino (2008).
\newblock Variable selection with incomplete covariate data.
\newblock {\em Biometrics\/}~{\em 64}, 1062--1069.

\bibitem[\protect\citeauthoryear{Claeskens and Hjort}{Claeskens and
  Hjort}{2008}]{ClaeskensHjort2008}
Claeskens, G. and N.~L. Hjort (2008).
\newblock {\em Model selection and model averaging}.
\newblock Cambridge: Cambridge University Press.

\bibitem[\protect\citeauthoryear{Cox}{Cox}{1972}]{Cox1972}
Cox, D.~R. (1972).
\newblock Regression models and life-tables.
\newblock {\em Journal of the Royal Statistical Society. Series B
  (Methodological)\/}~{\em 34\/}(2), 187--220.

\bibitem[\protect\citeauthoryear{Crevecoeur, Antonio, and Verbelen}{Crevecoeur
  et~al.}{2019}]{crevecoeur2019}
Crevecoeur, J., K.~Antonio, and R.~Verbelen (2019).
\newblock Modeling the number of hidden events subject to observation delay.
\newblock {\em European Journal of Operational Research\/}~{\em 277\/}(3), 930
  -- 944.

\bibitem[\protect\citeauthoryear{Dempster, Laird, and Rubin}{Dempster
  et~al.}{1977}]{dempster1977maximum}
Dempster, A.~P., N.~M. Laird, and D.~B. Rubin (1977).
\newblock Maximum likelihood from incomplete data via the {EM} algorithm.
\newblock {\em Journal of the Royal Statistical Society. Series B
  (Methodological)\/}~{\em 39\/}(1), 1--38.

\bibitem[\protect\citeauthoryear{Dirick, Claeskens, and Baesens}{Dirick
  et~al.}{2015}]{DirickClaeskensBaesens2015}
Dirick, L., G.~Claeskens, and B.~Baesens (2015).
\newblock An {A}kaike information criterion for multiple event mixture cure
  models.
\newblock {\em European Journal of Operational Research\/}~{\em 241}, 449--457.

\bibitem[\protect\citeauthoryear{Donker, van Boven, van Ballegooijen, van't
  Klooster, Wielders, and Wallinga}{Donker et~al.}{2011}]{Donker}
Donker, T., M.~van Boven, W.~van Ballegooijen, T.~van't Klooster, C.~Wielders,
  and J.~Wallinga (2011).
\newblock Nowcasting pandemic influenza {A/H1N1} 2009 hospitalizations in the
  {N}etherlands.
\newblock {\em European Journal of Epidemiology\/}~{\em 26}, 195--201.

\bibitem[\protect\citeauthoryear{England and Verrall}{England and
  Verrall}{2002}]{EnglandVerrall2002}
England, P.~D. and R.~J. Verrall (2002).
\newblock Stochastic claims reserving in general insurance.
\newblock {\em British Actuarial Journal\/}~{\em 8\/}(3), 443--518.

\bibitem[\protect\citeauthoryear{Farrington, Andrews, Beale, and
  Catchpole}{Farrington et~al.}{1996}]{Farrington1996}
Farrington, C.~P., N.~J. Andrews, A.~D. Beale, and M.~A. Catchpole (1996).
\newblock A statistical algorithm for the early detection of outbreaks of
  infectious disease.
\newblock {\em Journal of the Royal Statistical Society: Series A (Statistics
  in Society)\/}~{\em 159\/}(3), 547--563.

\bibitem[\protect\citeauthoryear{Günther, Bender, Katz, Küchenhoff, and
  Höhle}{Günther et~al.}{2021}]{Gunther2021}
Günther, F., A.~Bender, K.~Katz, H.~Küchenhoff, and M.~Höhle (2021).
\newblock Nowcasting the \textsc{COVID}-19 pandemic in \textsc{B}avaria.
\newblock {\em Biometrical Journal\/}~{\em 63\/}(3), 490--502.

\bibitem[\protect\citeauthoryear{Greene, McGough, Culp, Graf, Lipsitch,
  Menzies, and Kahn}{Greene et~al.}{2021}]{Greene2021}
Greene, S.~K., S.~F. McGough, G.~M. Culp, L.~E. Graf, M.~Lipsitch, N.~A.
  Menzies, and R.~Kahn (2021, Jan).
\newblock Nowcasting for real-time \textsc{COVID}-19 tracking in \textsc{N}ew
  \textsc{Y}ork \textsc{C}ity: An evaluation using reportable disease data from
  early in the pandemic.
\newblock {\em JMIR Public Health Surveill\/}~{\em 7\/}(1), e25538.

\bibitem[\protect\citeauthoryear{Haastrup and Arjas}{Haastrup and
  Arjas}{1996}]{Haastrup}
Haastrup, S. and E.~Arjas (1996).
\newblock Claims reserving in continuous time: a nonparametric {B}ayesian
  approach.
\newblock {\em {ASTIN} {B}ulletin\/}~{\em 26}, 139--164.

\bibitem[\protect\citeauthoryear{Hachemeister and Stanard}{Hachemeister and
  Stanard}{1975}]{hachemeister1975ibnr}
Hachemeister, C.~A. and J.~N. Stanard (1975).
\newblock {IBNR} claims count estimation with static lag functions.
\newblock In {\em Spring Meeting of the Casualty Actuarial Society}.

\bibitem[\protect\citeauthoryear{Harris}{Harris}{1990}]{harris1990reporting}
Harris, J.~E. (1990).
\newblock Reporting delays and the incidence of {AIDS}.
\newblock {\em Journal of the American Statistical Association\/}~{\em
  85\/}(412), 915--924.

\bibitem[\protect\citeauthoryear{H{ö}hle and an~der Heiden}{H{ö}hle and
  an~der Heiden}{2014}]{Hohle2014}
H{ö}hle, M. and M.~an~der Heiden (2014).
\newblock Bayesian nowcasting during the {STEC O104:H4} outbreak in {G}ermany,
  2011.
\newblock {\em Biometrics\/}~{\em 70\/}(4), 993--1002.

\bibitem[\protect\citeauthoryear{Hiabu, Mammen, Martínez-Miranda, and
  Nielsen}{Hiabu et~al.}{2021}]{Hiabu2020}
Hiabu, M., E.~Mammen, M.~D. Martínez-Miranda, and J.~P. Nielsen (2021).
\newblock Smooth backfitting of proportional hazards with multiplicative
  components.
\newblock {\em Journal of the American Statistical Association\/}.

\bibitem[\protect\citeauthoryear{Jewell}{Jewell}{1989}]{jewell1989}
Jewell, W.~S. (1989).
\newblock Predicting {IBNYR} events and delays: {I}. {C}ontinuous time.
\newblock {\em {ASTIN} {B}ulletin\/}~{\em 19}, 25--55.

\bibitem[\protect\citeauthoryear{Jung and Tremayne}{Jung and
  Tremayne}{2011}]{Jung2011}
Jung, R.~C. and A.~R. Tremayne (2011).
\newblock Useful models for time series of counts or simply wrong ones?
\newblock {\em AStA Advances in Statistical Analysis\/}~{\em 95\/}(1), 59--91.

\bibitem[\protect\citeauthoryear{Kalbfleisch, Lawless, and
  Robinson}{Kalbfleisch et~al.}{1991}]{kalbfleisch1991methods}
Kalbfleisch, J., J.~Lawless, and J.~Robinson (1991).
\newblock Methods for the analysis and prediction of warranty claims.
\newblock {\em Technometrics\/}~{\em 33\/}(3), 273--285.

\bibitem[\protect\citeauthoryear{Kalbfleisch and Lawless}{Kalbfleisch and
  Lawless}{1989}]{Kalbfleish1989}
Kalbfleisch, J.~D. and J.~F. Lawless (1989).
\newblock Inference based on retrospective ascertainment: An analysis of the
  data on transfusion-related {AIDS}.
\newblock {\em Journal of the American Statistical Association\/}~{\em
  84\/}(406), 360--372.

\bibitem[\protect\citeauthoryear{Kalbfleisch and Lawless}{Kalbfleisch and
  Lawless}{1991}]{KalbfleischLawless1991}
Kalbfleisch, J.~D. and J.~F. Lawless (1991).
\newblock Regression models for right truncated data with applications to
  {AIDS} incubation times and reporting lags.
\newblock {\em Statistica Sinica\/}~{\em 1\/}(1), 19--32.

\bibitem[\protect\citeauthoryear{Lagakos, Barraj, and Gruttola}{Lagakos
  et~al.}{1988}]{lagakos1988}
Lagakos, S.~W., L.~M. Barraj, and V.~D. Gruttola (1988).
\newblock Nonparametric analysis of truncated survival data, with application
  to {AIDS}.
\newblock {\em Biometrika\/}~{\em 75\/}(3), 515--523.

\bibitem[\protect\citeauthoryear{Lawless}{Lawless}{1994}]{Lawless1994}
Lawless, J. (1994).
\newblock Adjustments for reporting delays and the prediction of occurred but
  not reported events.
\newblock {\em Canadian Journal of Statistics\/}~{\em 22\/}(1), 15--31.

\bibitem[\protect\citeauthoryear{Louis}{Louis}{1982}]{Louis1982}
Louis, T.~A. (1982).
\newblock Finding the observed information matrix when using the {EM}
  algorithm.
\newblock {\em Journal of the Royal Statistical Society. Series B
  (Methodological)\/}~{\em 44\/}(2), 226--233.

\bibitem[\protect\citeauthoryear{Mack}{Mack}{1991}]{mack1991simple}
Mack, T. (1991).
\newblock A simple parametric model for rating automobile insurance or
  estimating {IBNR} claims reserves.
\newblock {\em ASTIN Bulletin\/}~{\em 21\/}(1), 93--109.

\bibitem[\protect\citeauthoryear{Mack}{Mack}{1993}]{mack1993distribution}
Mack, T. (1993).
\newblock Distribution-free calculation of the standard error of chain ladder
  reserve estimates.
\newblock {\em ASTIN {B}ulletin\/}~{\em 23\/}(02), 213--225.

\bibitem[\protect\citeauthoryear{Mart{\'{i}}nez~Miranda, Nielsen, Sperlich, and
  Verrall}{Mart{\'{i}}nez~Miranda et~al.}{2013}]{Martinez2013}
Mart{\'{i}}nez~Miranda, M.~D., J.~P. Nielsen, S.~Sperlich, and R.~Verrall
  (2013).
\newblock Continuous chain ladder: Reformulating and generalizing a classical
  insurance problem.
\newblock {\em Expert Systems with Applications\/}~{\em 40\/}(14), 5588 --
  5603.

\bibitem[\protect\citeauthoryear{McGough, Johansson, Lipsitch, and
  Menzies}{McGough et~al.}{2020}]{McGough2020}
McGough, S.~F., M.~A. Johansson, M.~Lipsitch, and N.~A. Menzies (2020).
\newblock Nowcasting by \textsc{B}ayesian smoothing: A flexible, generalizable
  model for real-time epidemic tracking.
\newblock {\em \textsc{PLOS} Computational Biology\/}~{\em 16\/}(4), 1--20.

\bibitem[\protect\citeauthoryear{Meng and Rubin}{Meng and
  Rubin}{1991}]{MengRubin1991}
Meng, X.-L. and D.~B. Rubin (1991).
\newblock Using {EM} to obtain asymptotic variance-covariance matrices: The
  {SEM} algorithm.
\newblock {\em Journal of the American Statistical Association\/}~{\em
  86\/}(416), 899--909.

\bibitem[\protect\citeauthoryear{Norberg}{Norberg}{1993}]{Norberg1993}
Norberg, R. (1993).
\newblock Prediction of outstanding liabilities in non-life insurance.
\newblock {\em ASTIN {B}ulletin\/}~{\em 23\/}(1), 95--115.

\bibitem[\protect\citeauthoryear{Norberg}{Norberg}{1999}]{Norberg1999}
Norberg, R. (1999).
\newblock Prediction of outstanding liabilities {II}. {M}odel variations and
  extensions.
\newblock {\em ASTIN {B}ulletin\/}~{\em 29\/}(1), 5--27.

\bibitem[\protect\citeauthoryear{Noufaily, Farrington, Garthwaite, Enki,
  Andrews, and Charlett}{Noufaily et~al.}{2016}]{noufaily2016detection}
Noufaily, A., P.~Farrington, P.~Garthwaite, D.~G. Enki, N.~Andrews, and
  A.~Charlett (2016).
\newblock Detection of infectious disease outbreaks from laboratory data with
  reporting delays.
\newblock {\em Journal of the American Statistical Association\/}~{\em
  111\/}(514), 488--499.

\bibitem[\protect\citeauthoryear{Noufaily, Ghebremichael-Weldeselassie, Enki,
  Garthwaite, Andrews, Charlett, and Farrington}{Noufaily
  et~al.}{2015}]{noufaily2015modelling}
Noufaily, A., Y.~Ghebremichael-Weldeselassie, D.~G. Enki, P.~Garthwaite,
  N.~Andrews, A.~Charlett, and P.~Farrington (2015).
\newblock Modelling reporting delays for outbreak detection in infectious
  disease data.
\newblock {\em Journal of the Royal Statistical Society: Series A (Statistics
  in Society)\/}~{\em 178\/}(1), 205--222.

\bibitem[\protect\citeauthoryear{Oakes}{Oakes}{1999}]{Oakes1999}
Oakes, D. (1999).
\newblock Direct calculation of the information matrix via the {EM} algorithm.
\newblock {\em Journal of the Royal Statistical Society Series B-Statistical
  Methodology\/}~{\em 61\/}(2), 479--482.

\bibitem[\protect\citeauthoryear{Pagano, Tu, Gruttola, and MaWhinney}{Pagano
  et~al.}{1994}]{Pagano1994}
Pagano, M., X.~M. Tu, V.~D. Gruttola, and S.~MaWhinney (1994).
\newblock Regression analysis of censored and truncated data: Estimating
  reporting- delay distributions and aids incidence from surveillance data.
\newblock {\em Biometrics\/}~{\em 50\/}(4), 1203--1214.

\bibitem[\protect\citeauthoryear{Renshaw and Verrall}{Renshaw and
  Verrall}{1998}]{renshaw1998stochastic}
Renshaw, A.~E. and R.~J. Verrall (1998).
\newblock A stochastic model underlying the chain-ladder technique.
\newblock {\em British Actuarial Journal\/}~{\em 4\/}(4), 903--923.

\bibitem[\protect\citeauthoryear{Schwarz}{Schwarz}{1978}]{Schwarz78}
Schwarz, G. (1978).
\newblock Estimating the dimension of a model.
\newblock {\em The Annals of Statistics\/}~{\em 6}, 461--464.

\bibitem[\protect\citeauthoryear{Sellero, Fern\'{a}ndez, Manteiga, Otero,
  Hervada, Fern\'{a}ndez, and Taboada}{Sellero et~al.}{1996}]{Sellero1996}
Sellero, C.~S., E.~V. Fern\'{a}ndez, W.~Manteiga, X.~L. Otero, X.~Hervada,
  E.~Fern\'{a}ndez, and X.~A. Taboada (1996).
\newblock Reporting delay: a review with simulation study and application to
  {S}panish data.
\newblock {\em Statistics in Medicine\/}~{\em 15\/}(3), 305--321.

\bibitem[\protect\citeauthoryear{Tabnak, M\"uller, Wang, Chiou, and Sun}{Tabnak
  et~al.}{2000}]{Tabnak2000}
Tabnak, F., H.~M\"uller, J.~Wang, J.-M. Chiou, and R.~Sun (2000).
\newblock A change-point model for reporting delays under change of
  \textsc{AIDS} case definition.
\newblock {\em European Journal of Epidemiology\/}~{\em 16}, 1135--1141.

\bibitem[\protect\citeauthoryear{Taylor}{Taylor}{2000}]{Taylor2000}
Taylor, G. (2000).
\newblock {\em Loss reserving: an actuarial perspective}.
\newblock Kluwer Academic Publishers.

\bibitem[\protect\citeauthoryear{van~de Kassteele, Eilers, and Wallinga}{van~de
  Kassteele et~al.}{2019}]{Kassteele2019}
van~de Kassteele, J., P.~Eilers, and J.~Wallinga (2019).
\newblock Nowcasting the number of new symptomatic cases during infectious
  disease outbreaks using constrained {P}-spline smoothing.
\newblock {\em Epidemiology\/}~{\em 30}, 737--745.

\bibitem[\protect\citeauthoryear{Verrall and W{\"u}thrich}{Verrall and
  W{\"u}thrich}{2016}]{verrall2016understanding}
Verrall, R.~J. and M.~V. W{\"u}thrich (2016).
\newblock Understanding reporting delay in general insurance.
\newblock {\em Risks\/}~{\em 4\/}(3), 25.

\bibitem[\protect\citeauthoryear{Wahl}{Wahl}{2019}]{Wahl2019}
Wahl, F. (2019).
\newblock Explicit moments for a class of micro-models in non-life insurance.
\newblock {\em Insurance: Mathematics and Economics\/}~{\em 89}, 140 -- 156.

\bibitem[\protect\citeauthoryear{Wu}{Wu}{2013}]{wu2013review}
Wu, S. (2013).
\newblock A review on coarse warranty data and analysis.
\newblock {\em Reliability Engineering \& System Safety\/}~{\em 114}, 1--11.

\bibitem[\protect\citeauthoryear{W{\"u}thrich and Merz}{W{\"u}thrich and
  Merz}{2008}]{WuthrichMerz2008}
W{\"u}thrich, M.~V. and M.~Merz (2008).
\newblock {\em Stochastic claims reserving methods in insurance}, Volume 435 of
  {\em Wiley Finance}.
\newblock John Wiley \& Sons.

\bibitem[\protect\citeauthoryear{W{\"u}thrich and Merz}{W{\"u}thrich and
  Merz}{2015}]{WuthrichMerz2015}
W{\"u}thrich, M.~V. and M.~Merz (2015).
\newblock {\em Stochastic Claims Reserving Manual: Advances in Dynamic
  Modeling}.
\newblock Swiss Finance Institute Research Paper No. 15--34.

\bibitem[\protect\citeauthoryear{Zhu, Lee, Wei, and Zhou}{Zhu
  et~al.}{2001}]{ZhuLeeWeiZhou2001}
Zhu, H., S.~Lee, B.~Wei, and J.~Zhou (2001).
\newblock Case‐deletion measures for models with incomplete data.
\newblock {\em Biometrika\/}~{\em 88\/}(3), 727--737.

\end{thebibliography}

\newpage

\appendix

\title{Supplementary material for \\ "Modeling the occurrence of events subject to a reporting delay via an EM algorithm"}
\date{\today}
\maketitle

This Appendix collects additional figures and tables regarding the case study on insurance nowcasting, discussed in Section~\ref{sec:case_study} of our paper entitled \textit{Modeling the occurrence of events subject to a reporting delay via an EM algorithm}.

\section{Case study}\label{app:add_case_study}

We analyze a data set with the occurrence and reporting dates of claims in a portfolio of general liability insurance policies for private individuals from a European insurance company. The data set used in this case study has also been studied in e.g.~\cite{AntonioPlat2014} with a stochastic model for the development of claims after reporting and \cite{crevecoeur2019} who also nowcast the IBNR claims, but exclusively put focus on the dynamics in the reporting process.

\subsection{Description of the insurance data set}\label{app:descr_plots}

Exposure is available in this dataset by month from January 2000 onwards, expressed as \textit{earned exposure}, i.e.~the fraction of policies actually exposed to risk during the period. This means that a policy providing insurance coverage only during 10 days in January will contribute 10/365th to the exposure of that month. Earned exposure is not available on a daily level so instead we transform the monthly exposure to daily exposure by dividing by the number of days in each month. This accounts for the varying month lengths.

\begin{figure}[ht!]%
\centering
\includegraphics[width=0.8\columnwidth]{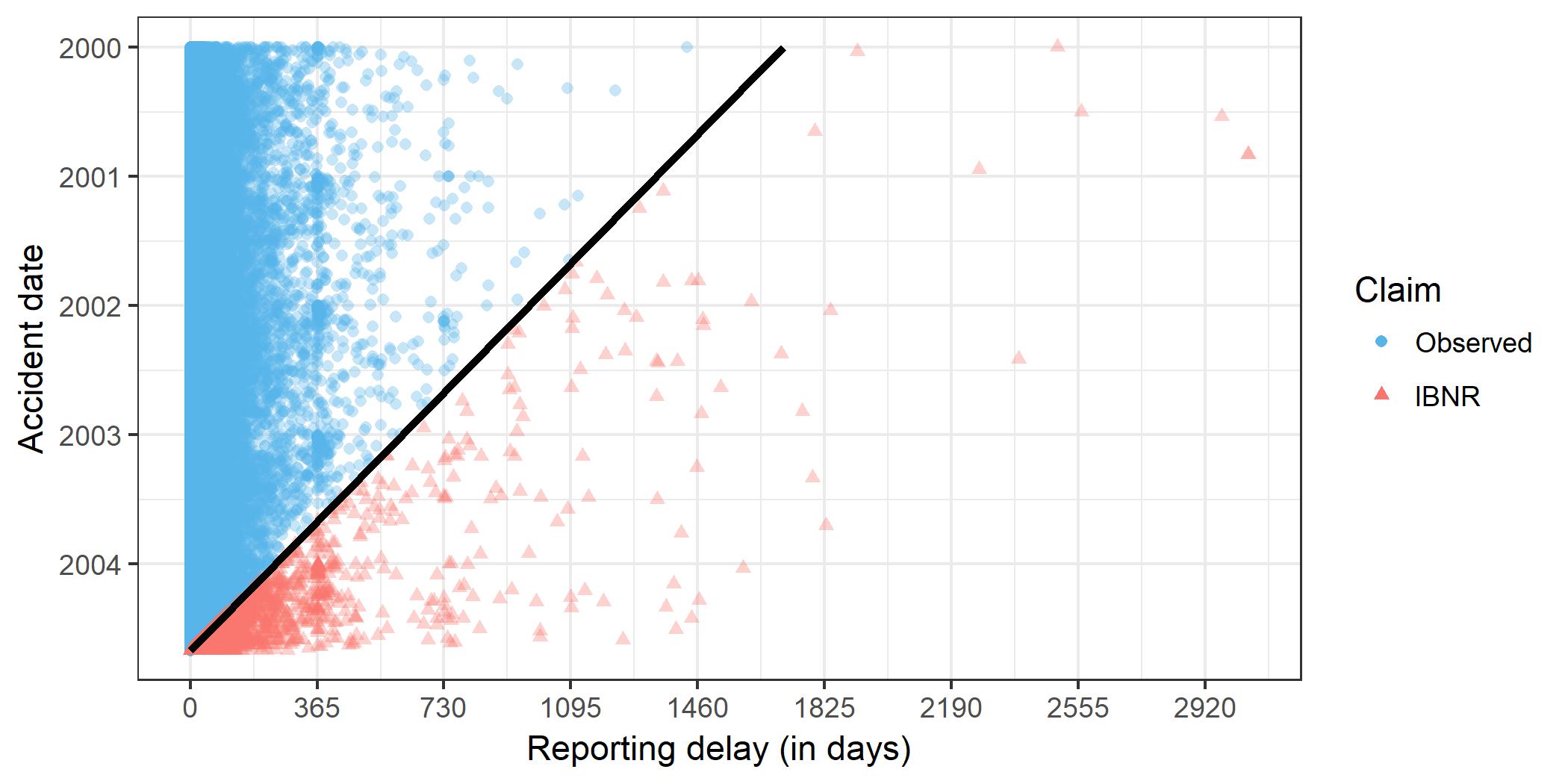}%
\caption{Daily run-off triangle of claims with occurrence dates between January 1, 2000 and August 31, 2004. The black line indicates the evaluation date, August 31, 2004. Only the claims in the upper triangle depicted as blue dots are observed at the evaluation date. The remaining claims in the lower triangle depicted as red triangles are the IBNR claims based on the full data set until August 2009 and have to be predicted.}%
\label{fig:daily_triangle}%
\end{figure}

To enable out-of-time prediction, we restrict our analysis to claims that have occurred between January 1, 2000 and August 31, 2004. We then study the nowcasting problem using evaluation date, say $\tau^{\star}$, at the end of this time window, on August 31, 2004 and want to estimate the total incurred but not reported (IBNR) claim count, as well as the reporting dates of these IBNR claims.
Based on the full data set until August 2009, \num{176671} claims have occurred during the time window between January 1, 2000 and August 31, 2004.
Due to a reporting delay, only \num{174624} of these have been reported by the evaluation date, as depicted in blue in the daily run-off triangle in Figure \ref{fig:daily_triangle}. The remaining 2047 are IBNR claims, i.e.~claims which have occurred between January 2000 and August 2004 but have only been reported after the evaluation date and before the end of the observation period. These are graphically illustrated in red in Figure~\ref{fig:daily_triangle}.

\subsection{Parametric models for the occurrence and reporting processes}\label{app:par_occ_rep}

\begin{figure}[htb!]%
\centering
\begin{subfigure}[t]{\textwidth}
\includegraphics[width = \textwidth]{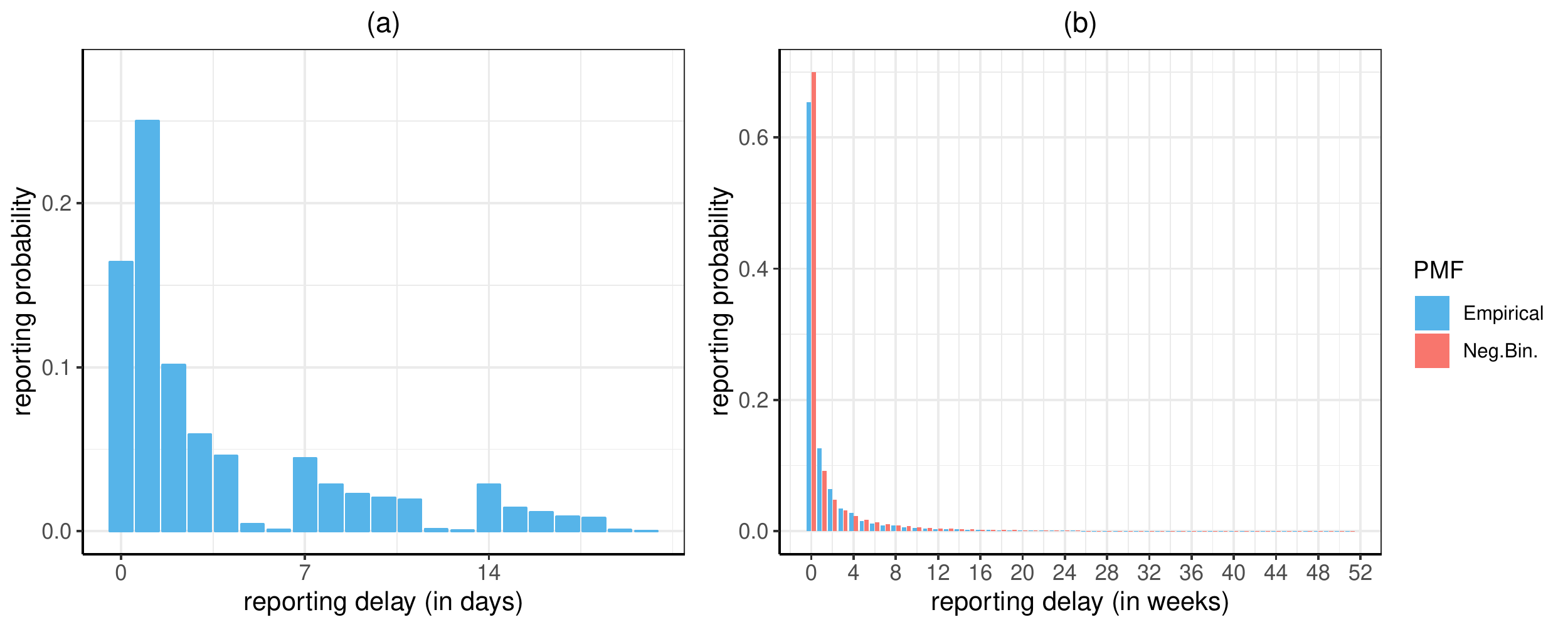}
\phantomcaption \label{fig:reporting_Monday}
\phantomcaption \label{fig:reporting_week}
\end{subfigure}
	%
	%
	%
\caption{(a) Bar plot of the empirical reporting delay distribution in the first 28 days for claims that occurred on a Monday. (b) Bar plot of the empirical reporting delay distribution in weeks and its negative binomial fit for the first year based on claims that occurred between January 2000 and August 2004 and have been reported before August 2009.
\label{fig:reporting}%
}
\end{figure}

Assisted by the data we specify a structure for the reporting probabilities $p_{td}(\bs{\theta},\bs{x}_{td})$ introduced in assumption~\ref{A2}. The barplot in Figure~\ref{fig:reporting_Monday} shows the empirical reporting probabilities in the first 28 days after occurrence for claims that occurred on a Monday. Reporting probabilities decrease over time and are low on Saturdays and Sundays. The claims pictured in Figure~\ref{fig:reporting_Monday} all occurred on a Monday. Therefore the first weekend corresponds to delays of 5 (Saturday) and 6 (Sunday) days. Following this intra-week pattern, we structure reporting delay as the product of week probabilities and day (or intra-week) probabilities:
\[ p_{td}(\bs{\theta},\bs{x}_{td}) = p^{\W}_{tw}(\bs{\theta},\bs{x}_{t})  \cdot  p^{\text{intra}}_{td}.\]
Here $p^{W}_{tw}$ denotes the probability of reporting an event from occurrence day $t$ in the $w$-th week after occurrence. The intra-week reporting probabilities are denoted with $p^{\text{intra}}_{td}$. The latter take values between 0 and 1 and sum to 1 over the days within the reporting week. Figure~\ref{fig:reporting_week} indicates that the negative binomial distribution is a good fit for the empirical week probabilities.

\subsection{Parameter estimates}\label{app:add_parm}

Complementary to the results shown in the paper, we give a detailed overview and discussion of the parameter estimates obtained for various model specifications.

\subsubsection{Occurrence and reporting processes: reporting week and intra-week probabilities}

We focus on the occurrence model \eqref{eq:occ} in combination with a reporting specified by the reporting week probabilities in \eqref{eq:mupW} and the intra-week reporting probabilities in \eqref{eq:P}. We report the estimated day probabilities from matrix $\P$ in Table~\ref{tab:P}. Only a small fraction of claims is being reported on Saturdays and nearly none on Sundays. In fact, in the entire observed part of the data, only 3 claims have been reported on Sunday.

\begin{table}[b!]
\centering
\caption{Maximum likelihood estimates of the day probabilities $\P$ within the reporting week. Separate reporting day probabilities are estimated for each day of the week (\textsf{dow}) of the occurrence date, as shown in the rows.}
\begin{tabular}{ llllllll }
\hline
 & \multicolumn{7}{c}{\textsf{wday}} \\ \cline{2-8}
\textsf{dow}          &			\textsf{wday1} 	&	\textsf{wday2} 	&	\textsf{wday3} 	   &	\textsf{wday4} 	&	\textsf{wday5} 	 &	\textsf{Saturday} 	& \textsf{Sunday} \\  \hline
\textsf{Monday} & 0.271 & 0.331 & 0.171 & 0.119 & 0.100 & 0.008 & 0.000 \\
\textsf{Tuesday} & 0.282 & 0.342 & 0.158 & 0.118  & 0.090  & 0.011 & 0.000 \\
\textsf{Wednesday} & 0.286 & 0.316 & 0.180  & 0.112 & 0.095 & 0.011 & 0.000 \\
\textsf{Thursday} & 0.278 & 0.337  & 0.156 & 0.114 & 0.097  & 0.019 & 0.000\\
\textsf{Friday}  & 0.303 & 0.264 & 0.160 & 0.120 & 0.096 & 0.057 & 0.000  \\
\textsf{Saturday}  & 0.389 & 0.211 & 0.148 & 0.109 & 0.097 & 0.046 & 0.000  \\
\textsf{Sunday}  & 0.407 & 0.222 & 0.157 & 0.109 & 0.096 & 0.009  & 0.000 \\
   \hline
\end{tabular}
 \label{tab:P}
\end{table}

\begin{figure}[h!]%
\centering
\includegraphics[width=\columnwidth]{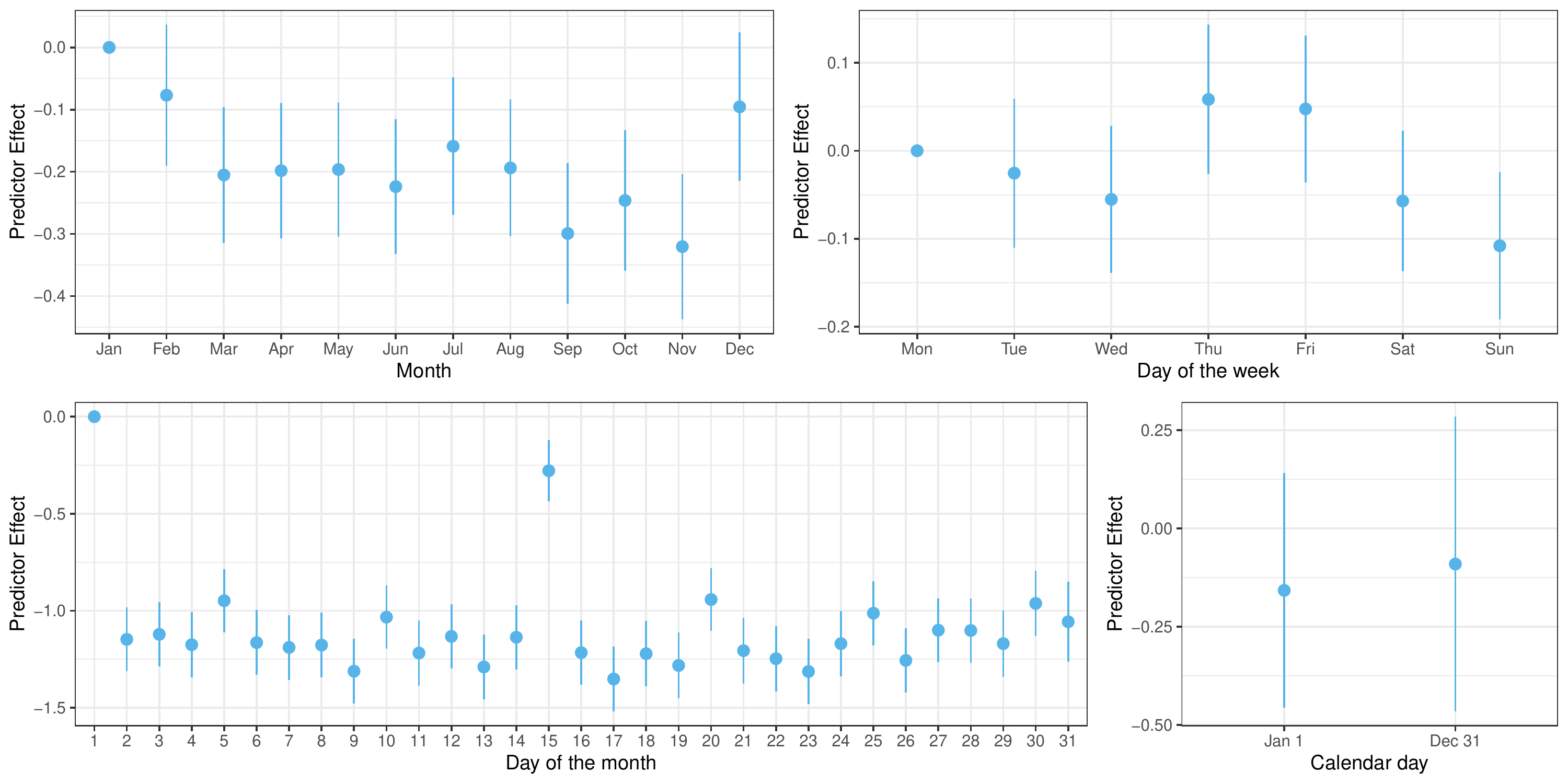}%
\caption{Maximum likelihood estimates and 95\% simultaneous confidence intervals for $\bs{\theta}$
corresponding to the categorical effects of the month, the day of the week and the day of the month of the occurrence date
in the negative binomial reporting delay model.}%
\label{fig:reportingModel}%
\end{figure}

The maximum likelihood estimates of the parameter vector $\bs{\alpha}$ from the claim occurrence model are shown in Figure~\ref{fig:occurrenceModel} in the paper. Next to these, the maximum likelihood estimates of the parameter vector $\bs{\theta}$ used in the negative binomial regression model of the reporting delay in weeks are displayed in Figure~\ref{fig:reportingModel}. All estimates are shown along with 95\% confidence intervals based on the inverse of the expected information matrix.
Simultaneous confidence intervals are constructed in these graphs using the Bonferroni correction to adjust for multiple comparisons.
For completeness, we also report that in the negative binomial model the intercept is estimated as 1.867 (with 95\% confidence interval $[1.717, 2.018]$) and the dispersion parameter $\phi$ as 0.177.

\subsection{Prediction of unreported claim counts}\label{app:pred_detail} 
\begin{figure}[htb!]%
\centering
 \begin{subfigure}[t]{\textwidth}
          \centering
					\caption{}%
					\vspace{-0.2cm}
					\includegraphics[width=\columnwidth]{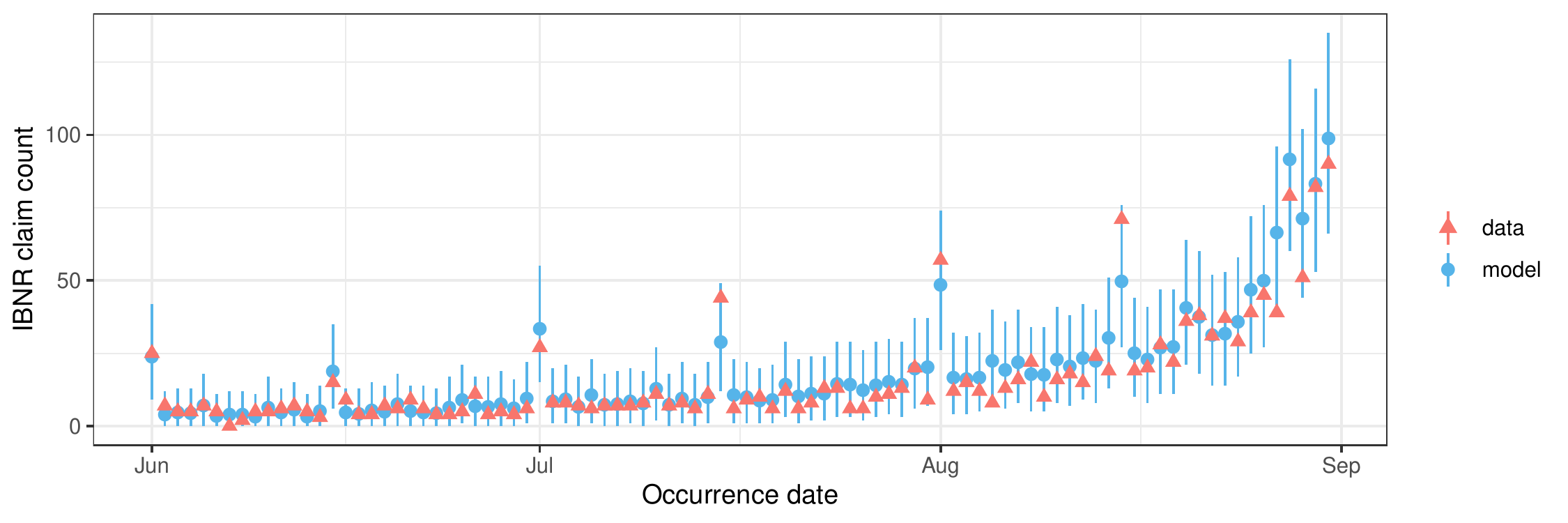}%
					\label{fig:IBNR_occurrence_day}%
					\vspace{-0.4cm}
     \end{subfigure}
		   \begin{subfigure}[t]{0.5\textwidth}
          \centering
					\caption{}%
					\vspace{-0.2cm}
					\includegraphics[width=\columnwidth]{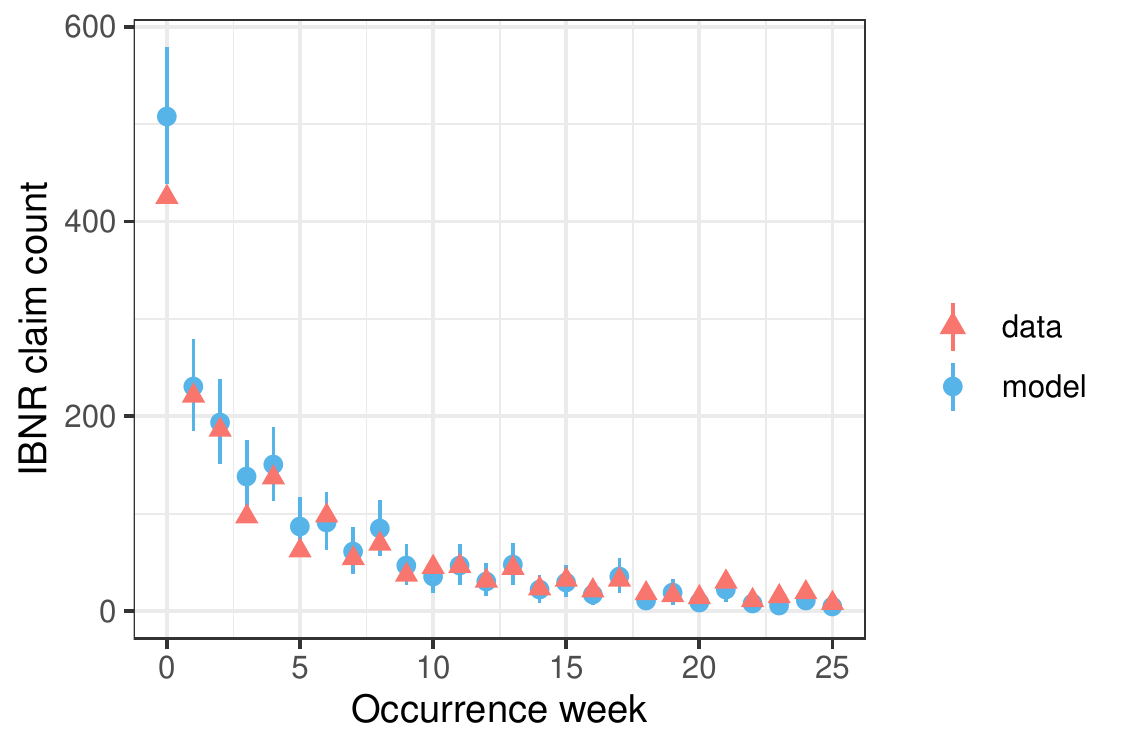}%
					\label{fig:IBNR_occurrence_week}%
     \end{subfigure}%
		 \begin{subfigure}[t]{0.5\textwidth}
          \centering
					\caption{}%
					\vspace{-0.2cm}
					\includegraphics[width=\columnwidth]{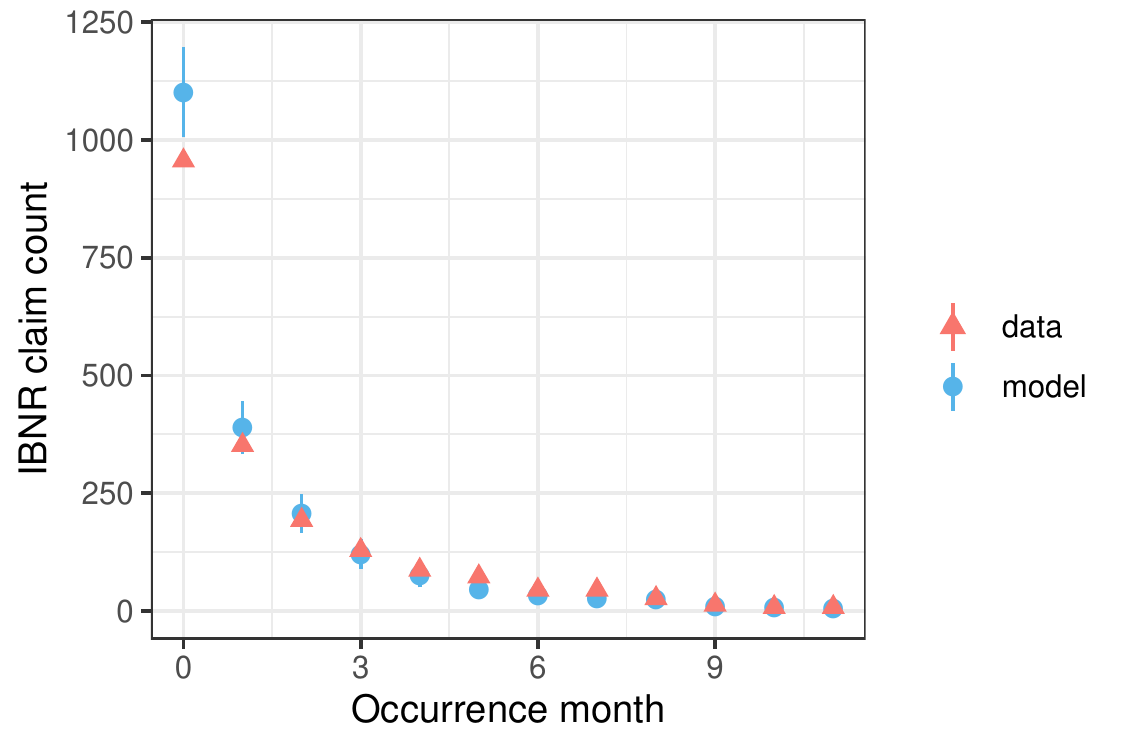}%
					\label{fig:IBNR_occurrence_month}%
     \end{subfigure}
\caption{Predictions of the IBNR claim counts and simultaneous 95\% prediction intervals by occurrence date. (a) Daily, for occurrence dates in between July 1 and August 31, 2004; (b) weekly (7 days) for the past 26 weeks; (c) monthly (30 days) for the past 12 months. %
\label{fig:IBNR_occurrence} }%
\end{figure}

We illustrate how the occurrence model \eqref{eq:occ} in combination with the reporting structure defined in \eqref{eq:mupW} and \eqref{eq:P} is used to forecast the number number of unreported claims for past occurrence dates. In Figure~\ref{fig:IBNR_occurrence_day} we plot point estimates and 95\% simultaneous prediction intervals for $N_t^{\IBNR}$ with $t$ corresponding to occurrences dates in between July 1, 2004, and August 31, 2004, i.e.~the last two months from our training period.  The predictions follow the same trend as the actual IBNR claim counts derived from the full data set until August 2009. We notice how IBNR claims are elevated on the first day and middle of each month, in line with our earlier findings.
In Figure \ref{fig:IBNR_occurrence_week} (resp.~Figure \ref{fig:IBNR_occurrence_month}) we group the occurrence dates by weeks (resp.~months) prior to the evaluation date and show the IBNR claim count predictions corresponding to the past 26 weeks (resp.~12 months).
We notice how, also over longer time spans, the predictions by occurrence week or month follow the pattern observed in the actual unreported counts.

\subsection{Modelling the intra week probabilities using a reverse time strategy}

\begin{figure}[ht!]%
\centering
\includegraphics[width=\columnwidth]{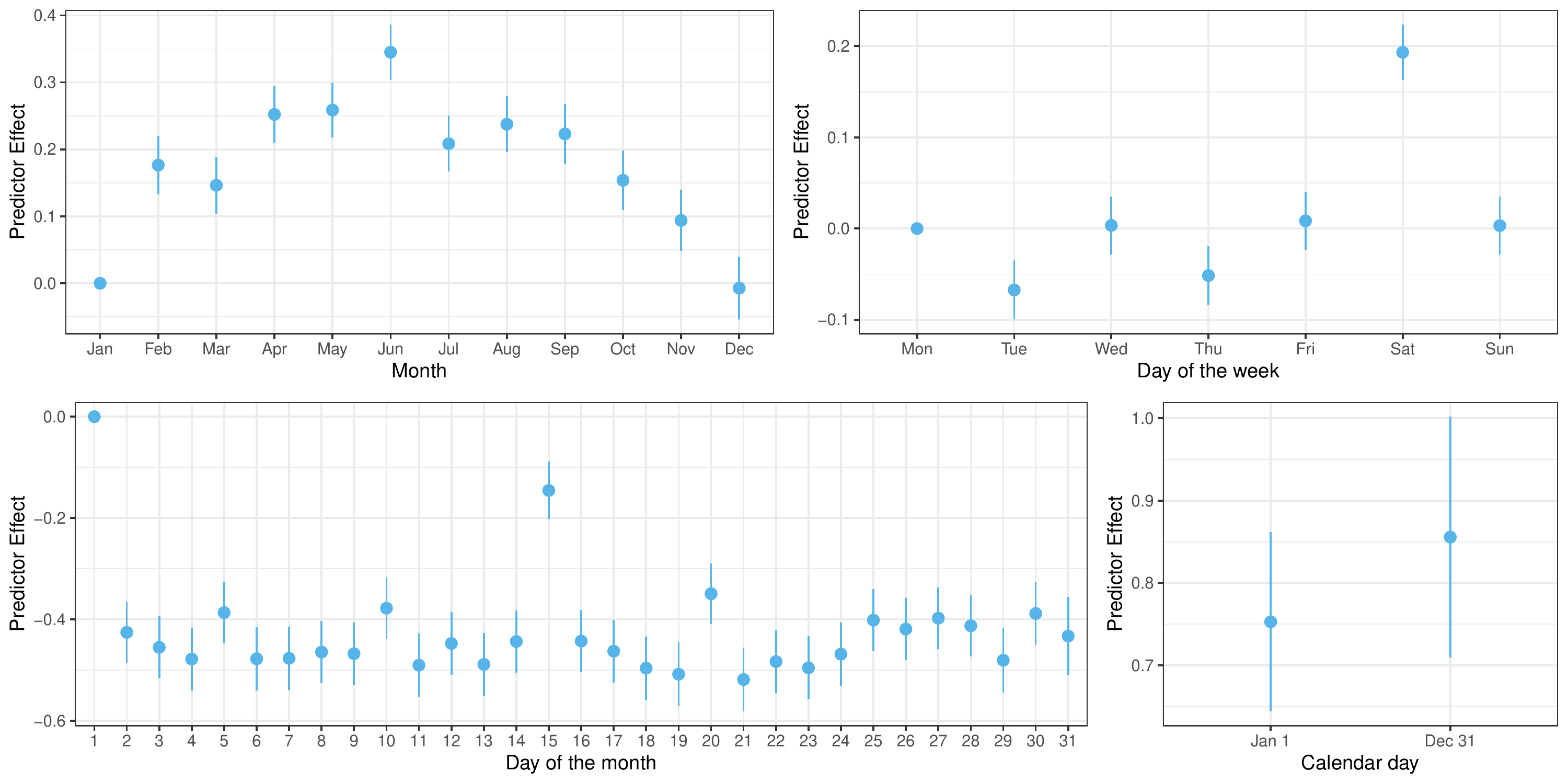}%
\caption{Maximum likelihood estimates for the parameters used in the occurrence model \eqref{eq:occ} jointly estimated with \eqref{eq:mupW} and \eqref{eq:q_probs}. 95\% simultaneous confidence intervals are constructed using the Bonferroni correction and the inverse of the expected information matrix.}%
\label{fig:occurrenceModel_holiday}%
\end{figure}

\begin{figure}[ht!]%
\centering
\includegraphics[width=\columnwidth]{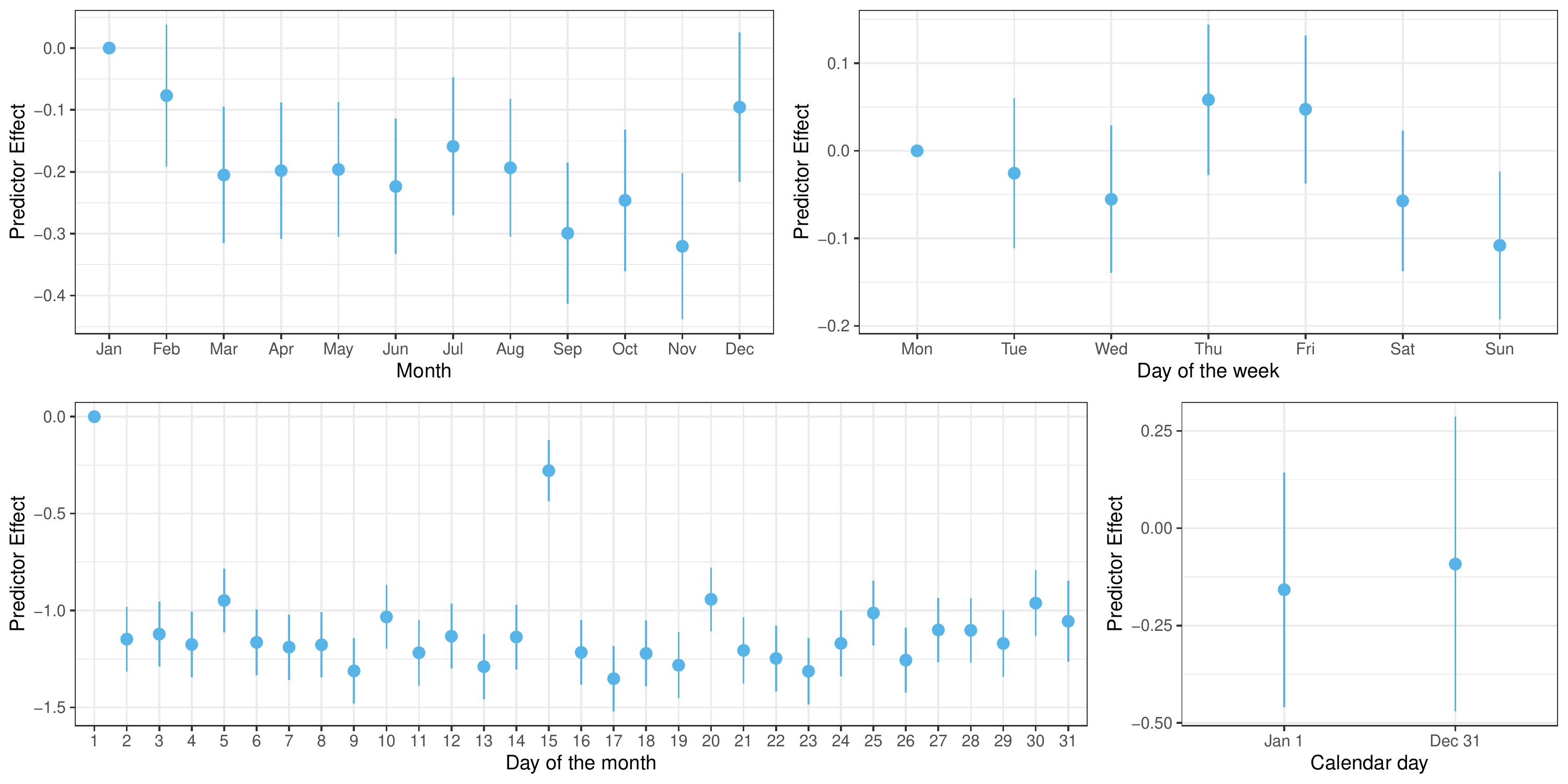}%
\caption{Maximum likelihood estimates for the parameters used in the reporting week model \eqref{eq:mupW} jointly estimated with \eqref{eq:occ} and \eqref{eq:q_probs}. 95\% simultaneous confidence intervals are constructed using the Bonferroni correction and the inverse of the expected information matrix.}%
\label{fig:reportingModel_holiday}%
\end{figure}

\begin{figure}[ht!]%
\centering
\includegraphics[width=\columnwidth]{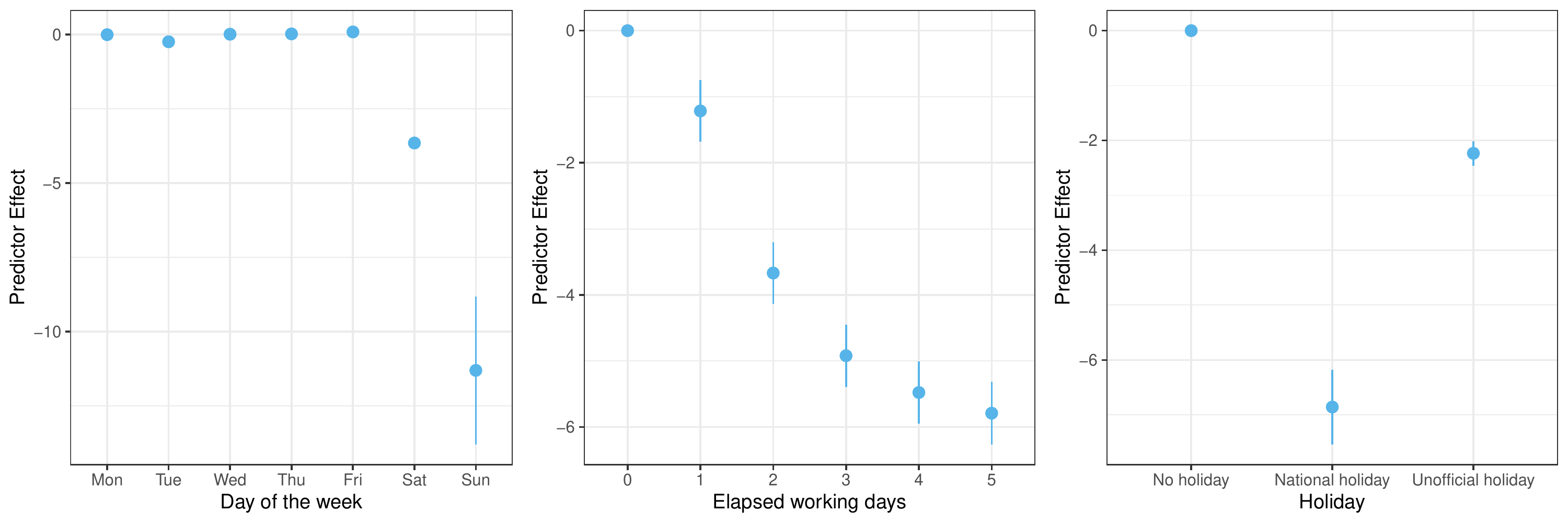}%
\caption{Maximum likelihood estimates for the parameters used in the intra-week reporting model \eqref{eq:q_probs} jointly estimated with \eqref{eq:occ} and \eqref{eq:mupW}. 95\% simultaneous confidence intervals are constructed using the Bonferroni correction and the inverse of the expected information matrix.}%
\label{fig:reportingModel_holiday_day}%
\end{figure}

We consider the occurrence model specified in \eqref{eq:occ} in combination with a reporting delay distribution with reporting week probabilities in \eqref{eq:mupW}
and intra-week probabilities obtained via the reverse time strategy in the presence of covariates, as specified in \eqref{eq:qprob}. Figure~\ref{fig:occurrenceModel_holiday} shows the estimates for the occurrence model in \eqref{eq:occ}, the estimates for the reporting week probabilities are shown in Figure~\ref{fig:reportingModel_holiday} and the intra-week reporting probabilities obtained via the reverse time strategy are in Figure~\ref{fig:reportingModel_holiday_day}.

\subsection{Directly modelling the reporting intensity}

\begin{figure}[h!]%
\centering
\includegraphics[width=\columnwidth]{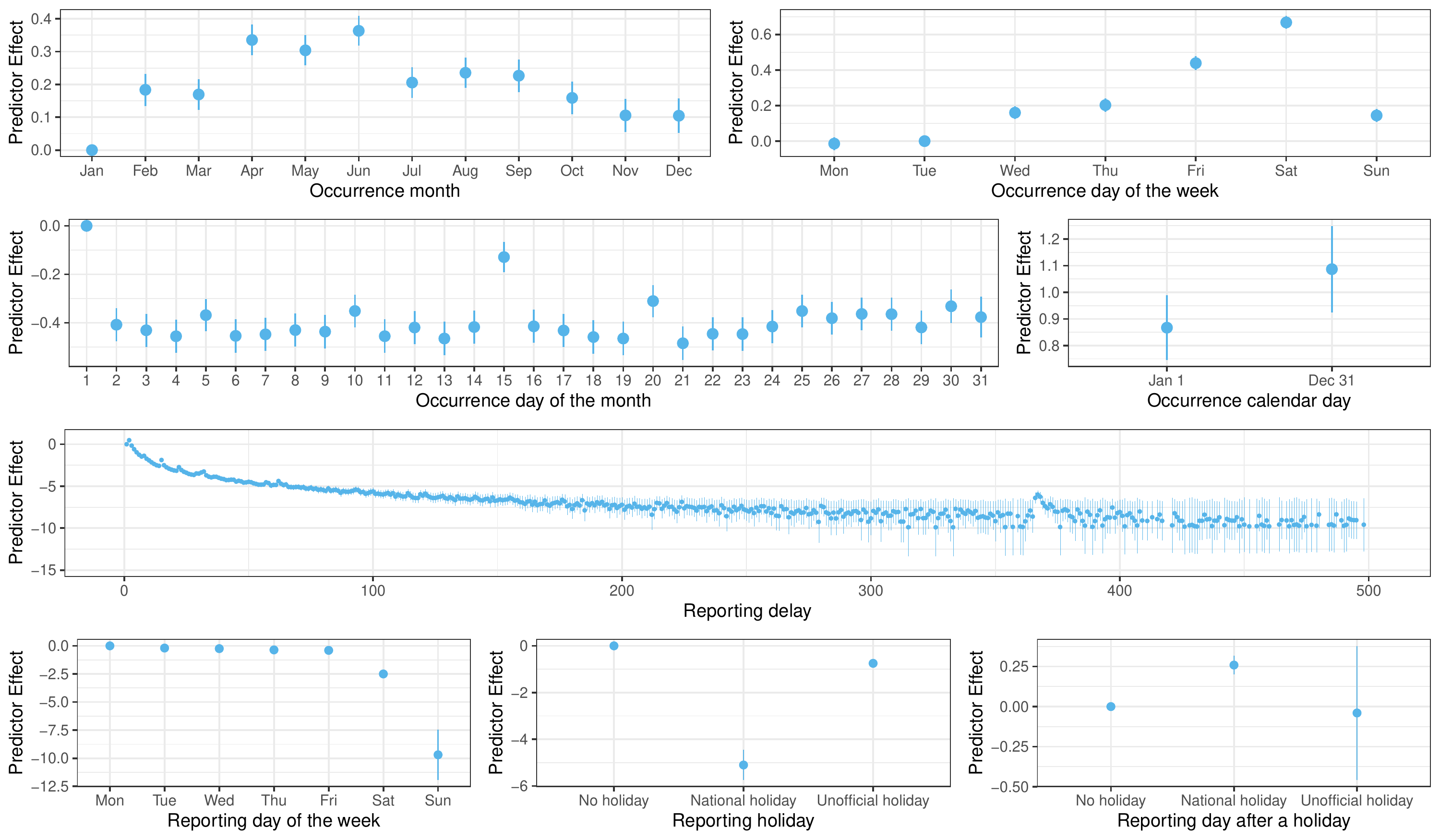}%
\caption{Maximum likelihood estimates for the parameters used in \eqref{eq:rep_int_1}. 95\% simultaneous confidence intervals are constructed using the Bonferroni correction and the inverse of the expected information matrix.}%
\label{fig:GLM_1}%
\end{figure}

\begin{figure}[h!]%
\centering
\includegraphics[width=\columnwidth]{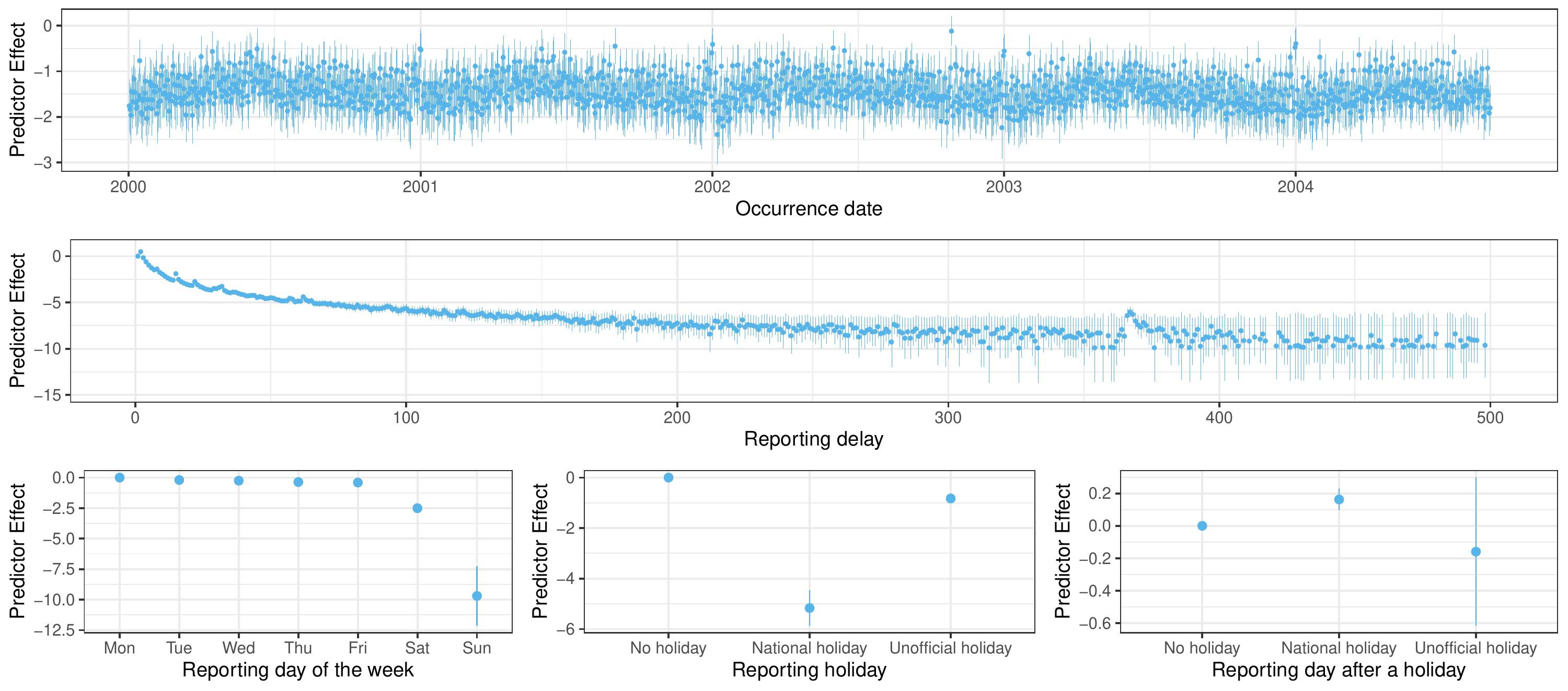}%
\caption{Maximum likelihood estimates for the parameters used in \eqref{eq:rep_int_2}. 95\% simultaneous confidence intervals are constructed using the Bonferroni correction and the inverse of the expected information matrix.}%
\label{fig:GLM_2}%
\end{figure}

We now consider two nowcasting models that directly structure the reporting intensity. These models are discussed in Section~\ref{sec:discrete_regr} of the paper, where the $N_{td}$ are independent Poisson distributed random variables with mean $\lambda_{td}$. Figure~\ref{fig:GLM_1} shows the parameter estimates for the specification from \eqref{eq:rep_int_1} and Figure~\ref{fig:GLM_2} displays the parameter estimates for the regression structure in \eqref{eq:rep_int_2}.

\subsection{Comparing nowcasting models with a moving window evaluation}

\begin{figure}[htb!]%
\centering

	\begin{subfigure}[t]{0.5\textwidth}
          \centering
					\caption{}%
					\vspace{-0.2cm}
					\includegraphics[width=\columnwidth]{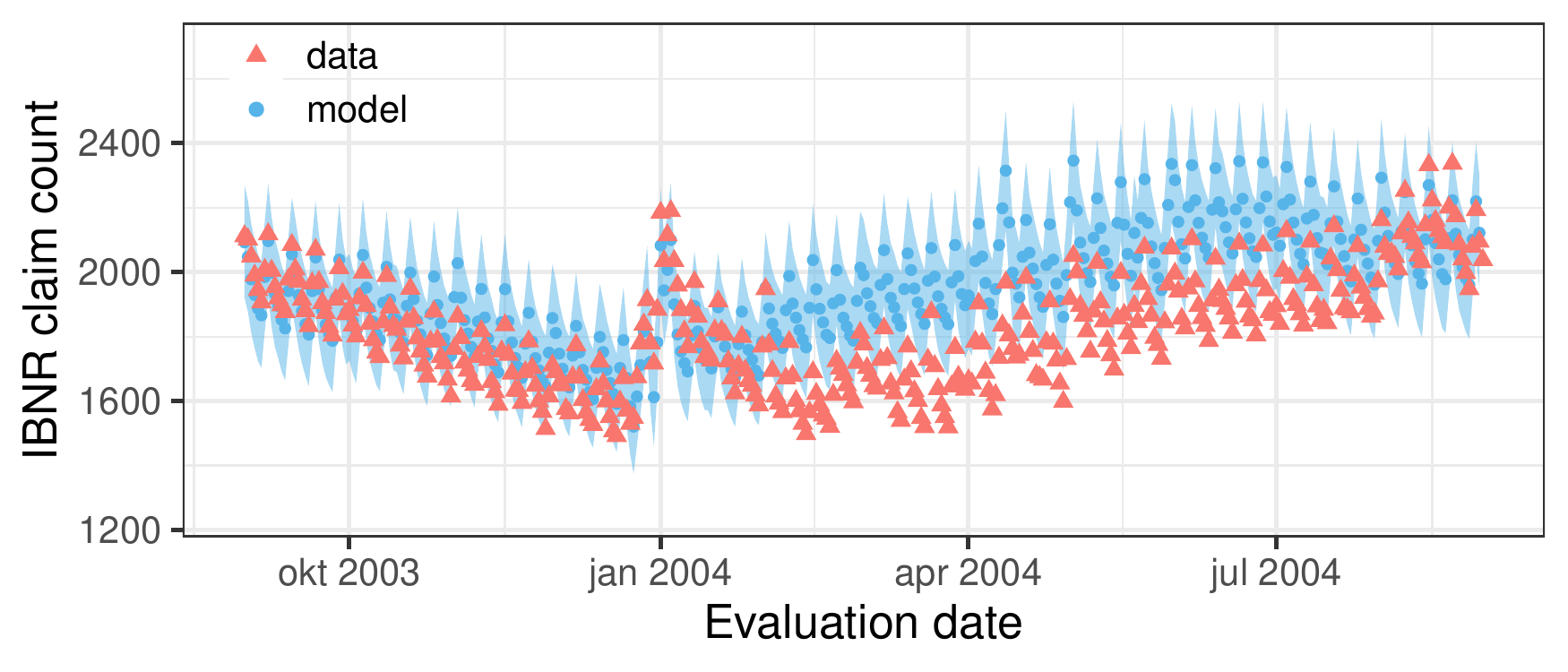}%
					\label{fig:ibnr_evolution_holiday_full}%
     \end{subfigure}%
		 \begin{subfigure}[t]{0.5\textwidth}
          \centering
					\caption{}%
					\vspace{-0.2cm}
					\includegraphics[width=\columnwidth]{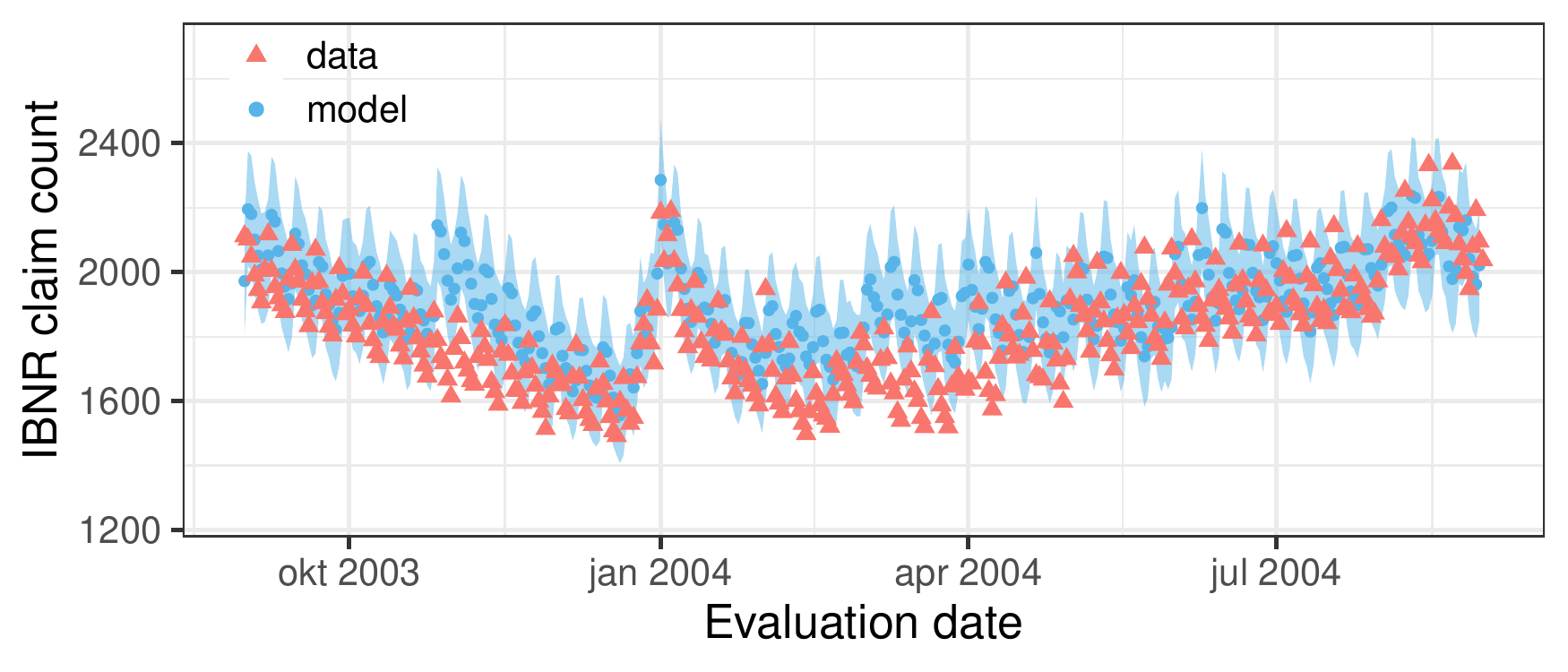}%
					\label{fig:ibnr_evolution_cl_full}%
     \end{subfigure}

     \begin{subfigure}[t]{0.5\textwidth}
          \centering
					\caption{}%
					\vspace{-0.2cm}
					\includegraphics[width=\columnwidth]{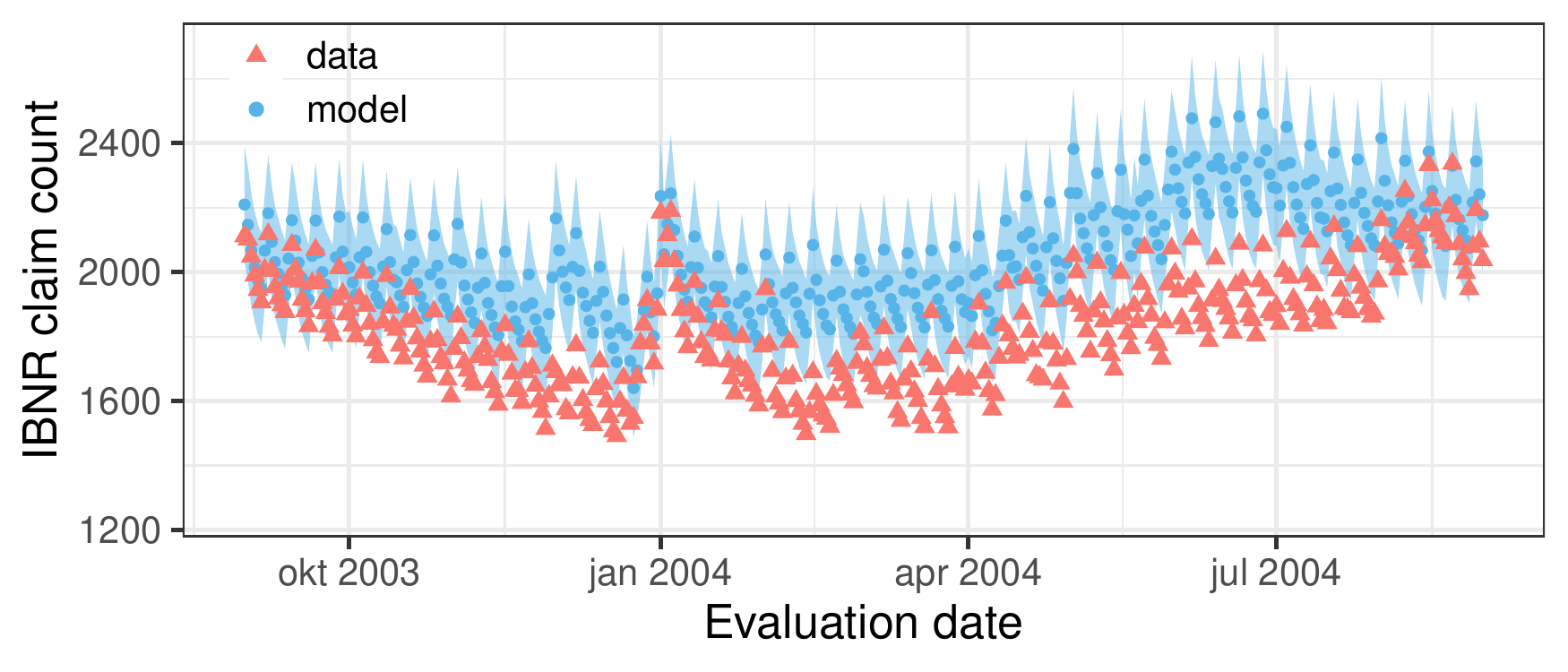}%
					\label{fig:ibnr_evolution_nowcasting_full}%
     \end{subfigure}%
		 \begin{subfigure}[t]{0.5\textwidth}
          \centering
					\caption{}%
					\vspace{-0.2cm}
					\includegraphics[width=\columnwidth]{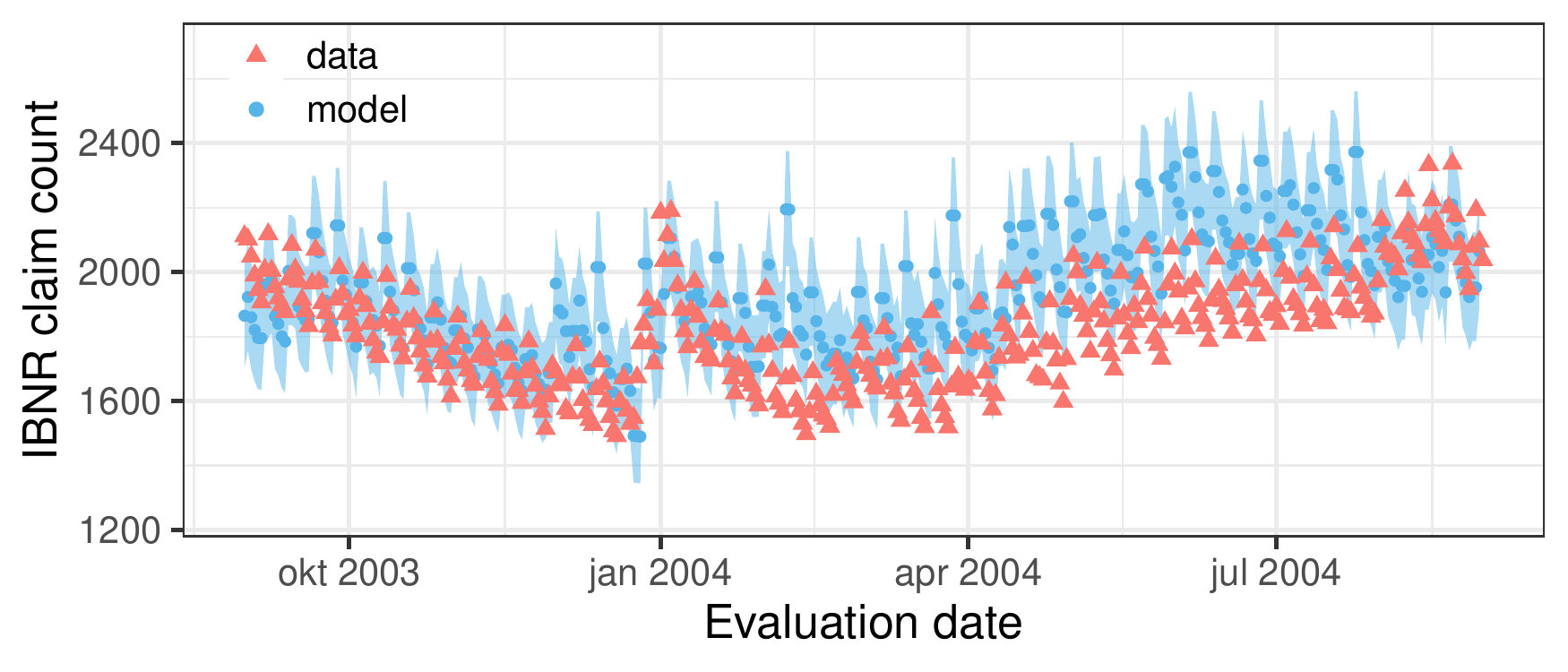}%
					\label{fig:ibnr_evolution_nowcasting_occ_full}%
     \end{subfigure}


\caption{Predictions and 95\% simultaneous prediction intervals of the total IBNR claim counts for varying evaluation dates from August 31, 2003 to August 31, 2004. (a) uses \eqref{eq:occ}, \eqref{eq:mupW} and \eqref{eq:q_probs}, (b) yearly chain ladder method \eqref{eq:chainladder_yearly}, (c) direct specification of reporting intensity in \eqref{eq:rep_int_1}, (d) direct specification of reporting intensity in \eqref{eq:rep_int_2}.
Red triangles: data, blue circles: model.}%
\vspace{-0.05cm}
\label{fig:IBNR_compared-full}%
\end{figure}

In addition to the results shown in Figure~\ref{fig:IBNR_compared-zoom}
in the paper 
(with evaluation dates between November 15 to February 15, 2004), we show the estimates of the total IBNR claim count with moving evaluation dates between August 31, 2003, and August 31, 2004.
 Figure~\ref{fig:ibnr_evolution_holiday_full} shows predictions obtained with the occurrence specification in \eqref{eq:occ} and the reporting model in \eqref{eq:mupW} and \eqref{eq:q_probs}.
Results obtained with the yearly chain ladder method are displayed in Figure~\ref{fig:ibnr_evolution_cl_full}. Panel \ref{fig:ibnr_evolution_nowcasting_full} shows the estimates for the IBNR claim counts obtained with a nowcasting model that directly specifies the reporting intensity along \eqref{eq:rep_int_1}, whereas the model leading to panel \ref{fig:ibnr_evolution_nowcasting_occ_full} uses \eqref{eq:rep_int_2}.

\end{document}